\documentclass[12pt,letterpaper]{article}
\usepackage[utf8]{inputenc}

\usepackage{color}
\definecolor{darkblue}{rgb}{0.1,0.1,.7}
\usepackage[colorlinks, linkcolor=darkblue, citecolor=darkblue, urlcolor=darkblue, linktocpage]{hyperref}
\usepackage[]{amsmath}
\usepackage[]{graphicx}
\usepackage[]{latexsym}
\usepackage[utf8]{inputenc}
\usepackage{slashed,graphicx,color,amsmath,amssymb}
\usepackage[mathscr]{eucal}
\usepackage{mathrsfs}
\usepackage[margin=10pt,font=small,labelfont=bf]{caption}
\usepackage{amscd}
\usepackage{bm}
\usepackage{xcolor}
\usepackage{upgreek}
\usepackage[square, comma, sort&compress,numbers]{natbib}
\usepackage[all,cmtip]{xy}
\usepackage[margin=1.7in]{geometry}
\usepackage{cleveref}
\usepackage[color=cyan!30!white,linecolor=red,textsize=footnotesize]{todonotes}
\usepackage{soul}
\usepackage{tikz}
\usepackage{pgflibraryarrows}
\usepackage{pgflibrarysnakes}
\usepackage{pgfplots}
\pgfplotsset{compat=1.8}
\tikzset{elegant/.style={smooth,thick,samples=50,magenta}}
\numberwithin{equation}{section}
 \makeatletter
 \g@addto@macro\bfseries{\boldmath}
\makeatletter
\if@todonotes@disabled

\else

\fi
\makeatother

\usepackage{amsmath}
\begin{document}	
\vspace*{-.3in} \thispagestyle{empty}
\begin{flushright}
\end{flushright}
{\Large
\begin{center}
{\bf Surface charges in Chern-Simons gravity with $T\bar{T}$ deformation}
\end{center}}
\renewcommand{\thefootnote}{\fnsymbol{footnote}}
\begin{center}
{Miao He$^{a,b,c}$, Song He$^{c,d}$\footnotemark[1], Yi-hong Gao$^{a,b}$\footnotemark[1]\footnotetext[1]{Corresponding author}}
\vspace{.2in}\\
\small{
\textit{$^{a}$School of Physical Sciences, University of Chinese Academy of Sciences, \\No.19A Yuquan Road, Beijing 100049, China}\\
\textit{$^{b}$CAS Key Laboratory of Theoretical Physics, Institute of Theoretical Physics, \\Chinese Academy of Sciences, Beijing 100190, China}
}\\
\textit{$^{c}$Center for Theoretical Physics and College of Physics, Jilin University, \\Changchun 130012, People's Republic of China}\\
\textit{$^{d}$Max Planck Institute for Gravitational Physics (Albert Einstein Institute), \\Am M\"uhlenberg 1, 14476 Golm, Germany}\\

E-mail:\ \ \begingroup\ttfamily\small
hemiao@itp.ac.cn,
hesong@jlu.edu.cn,
gaoyh@itp.ac.cn
\endgroup

%\vspace{.1in}

\end{center}
%\vspace{.5in}
\begin{abstract}
\normalsize{The $T\bar{T}$ deformed 2D CFTs correspond to AdS$_3$ gravity with Dirichlet boundary condition at finite cutoff or equivalently a mixed boundary condition at spatial infinity. In this work, we use the latter perspective and Chern-Simons formalism of AdS$_3$ gravity to construct the surface charges and associated algebra in $T\bar{T}$ deformed theories. Starting from the Ba\~nados geometry, we obtain the Chern-Simons gauge fields for the $T\bar{T}$ deformed geometry, which are parametrized by two independent charges. With help of the mixed boundary condition, the residual gauge symmetries of the deformed gauge fields and the associated surface charges were obtained respectively. The charge algebra turns out to be a non-linear deformed Virasoro algebra, which was obtained in different way by applying the cutoff perspective. Finally, we propose a way to construct the time-independent charges from these surface charges and they satisfy the field-dependent Virasoro algebra.}
\end{abstract}

%\begin{center}
%Keywords
%\end{center}

\newpage

\setcounter{page}{1}
\noindent\rule{\textwidth}{.1pt}\vspace{-1.2cm}
\begingroup
\hypersetup{linkcolor=black}
\tableofcontents
\endgroup
\noindent\rule{\textwidth}{.2pt}
\renewcommand{\thefootnote}{\arabic{footnote}}
%%%%%%%%%%%%%%%%%%%%%
%%%%%%%%%%%%%%%%%%%%%
%%%%%%%%%%%%%%%%%%%%%%%%%%%%%%%%%%
\section{Introduction}
\label{sec:1}
The $T\bar{T}$ deformed 2D CFTs attract a lot of interests because of its integrability and holographic duality~\cite{Smirnov:2016lqw, Cavaglia:2016oda, McGough:2016lol}. The action of $T\bar{T}$ deformation is defined by a flow triggered by the determinant of the stress tensor, which is also called $T\bar{T}$ operator. Although the deformation is irrelevant and the $T\bar{T}$ flow pushes a theory from IR to UV, many physical quantities can be expressed in terms of the un-deformed quantities, such as the Lagrangian~\cite{Bonelli:2018kik}, finite-size spectrum~\cite{Cavaglia:2016oda}, and partition function~\cite{Datta:2018thy, Aharony:2018bad}.  The $T\bar{T}$ deformation has many equivalent descriptions from various perspective. The $T\bar{T}$ deformation can be treated as the original theory dressed by the JT gravity~\cite{Dubovsky:2017cnj,Dubovsky:2018bmo,Ishii:2019uwk}. It can be also realized by a random coordinate transformation~\cite{Cardy:2018sdv}. Further, in \cite{Conti:2018tca,Conti:2019dxg}, $T\bar{T}$ deformed theories can be obtained by a specific field-dependent local change of coordinates in the undeformed theories. More recently, $T\bar{T}$ deformed theory can be also reformulated as non-critical string theory~\cite{Callebaut:2019omt,Tolley:2019nmm}.
\par
In AdS$_3$/CFT$_2$ context, it was proposed that the $T\bar{T}$ deformed CFT corresponds to the cutoff AdS$_3$ gravity at finite radial~\cite{McGough:2016lol,Kraus:2018xrn}, and the cutoff radius is related to the deformation parameter. The finite-size spectrum turned out to be the quasi-local energy of the BTZ black hole at finite radius. The $T\bar{T}$ flow equation coincides with the Hamilton-Jacobi equation governing the radial evolution of the classical gravity action in AdS$_3$~\cite{McGough:2016lol,Shyam:2017znq, Donnelly:2019pie}. It was known that the Dirichlet boundary conditions at finite radius correspond to the mixed boundary conditions at infinity\cite{Klebanov:1999tb,Witten:2001ua,Papadimitriou:2007sj}. An alternative holographic description is imposing a mixed boundary condition at the asymptotic AdS$_3$ boundary~\cite{Guica:2019nzm}. It turned out that the mixed boundary condition leads to a deformed bulk solution, which can be constructed by a field-dependent coordinate transformation~\cite{Guica:2019nzm}. AdS$_3$ gravity with the mixed boundary condition also reproduced the deformed spectrum. The boundary dynamics of AdS$_3$ with the mixed boundary condition can be described by the $T\bar{T}$ deformed coadjoint orbit of the Virasoro group~\cite{Ouyang:2020rpq,He:2020hhm}. Many holographic features of the $T\bar{T}$ deformed CFT have been explored~\cite{Giribet:2017imm,Donnelly:2018bef,Chen:2018eqk,Jeong:2019ylz,Grieninger:2019zts,Jafari:2019qns,Chen:2019mis,Mazenc:2019cfg,Li:2020pwa,Li:2020zjb,Caputa:2020lpa,Hirano:2020ppu,Araujo:2018rho,Babaei-Aghbolagh:2020kjg}. For more recent progresses, one can refer to the review of the $T\bar{T}$ deformation~\cite{Jiang:2019epa}.
\par
It has been shown that the $T\bar{T}$ deformation preserves integrability of the original theories~\cite{Smirnov:2016lqw,LeFloch:2019wlf}. Alternatively, it is interesting topic to investigate quantum integrable/chaotic signals of chaotical CFTs with $T\bar{T}$ deformation. The out of time ordered correlation function (OTOC) has been used to capture quantum integrability/chaos. As basic ingredients of OTOC, the correlation functions for $T\bar{T}$ deformation are also studied~\cite{Cardy:2019qao,He:2019vzf,He:2019ahx,Kruthoff:2020hsi,He:2020udl,He:2020cxp,He:2020qcs}. Recently, The $T\bar{T}$ deformations have been considered in other theories including integrable lattice models and non-relativistic integrable field theories~\cite{Cardy:2018jho,Marchetto:2019yyt,Cardy:2020olv,Medenjak:2020ppv,Jiang:2020nnb,Chen:2020jdi}. To understand the underline algebra/symmetry structure of the deformation in holography, the associated charges and their algebras have been explored on boundary field side and gravity side. The calculation from the boundary field side shows that some additional winding terms in Poisson brackets are not fixed due to certain ambiguities of the field-dependent coordinates~\cite{Jorjadze:2020ili,Guica:2020uhm}. On the gravity side, the charge algebra was obtained by considering 3D gravity with Dirichlet boundary conditions on a finite boundary~\cite{Guica:2019nzm,Kraus:2021cwf}. In AdS$_3$/CFT$_2$, the Chern-Simons formalism is a powerful tool to study the boundary dynamics and asymptotic symmetries~\cite{Brown:1986nw,Banados:1998gg,Banados:1998ta,Cotler:2018zff}. Even more, it can be naturally generalized to higher spin gravity~\cite{Henneaux:2010xg, Campoleoni:2010zq, Gutperle:2011kf}.

In the present work, we prefer to use the Chern-Simons form to study the asymptotic symmetries of AdS$_3$ with mixed boundary conditions. An analogy to the Ba\~nados geometry, we rewrite the deformed AdS$_3$ solution into Chern-Simons gauge fields, which are also parametrized by two independent functions. After imposing the mixed boundary condition, we find the residual gauge symmetries and associated surface charges in Chern-Simons theory. The resulting charge algebra turns out to be a non-linear deformation of the Virasoro algebra. The same charge algebra was also obtained by using the covariant phase space method~\cite{Kraus:2021cwf}. Further, we systematically construct the time-independent charges which satisfy the 
field-dependent Virasoro algebra. This result is in agreement with the conclusion in~\cite{Guica:2019nzm}.
\par
This paper is organized as follows: Section~\ref{sec:2} is a review of the global symmetries in Chern-Simons theory. In section~\ref{sec:3}, we rewrite the deformed solutions of AdS$_3$ in the Chern-Simons form. The $T\bar{T}$ deformed gauge field can be parametrized by two classes of independent deformed charges. In section~\ref{sec:4}, we obtain a set of residual gauge transformations which keep the deformed gauge connections asymptotically invariant. The residual gauge transformations generate a set of surface charges which satisfy the non-linear deformed Virasoro algebra. We comment on the surface charges and the charge algebra in section~\ref{sec:5}. Conclusions and discussions are given in section~\ref{sec:6}. Some calculation details are presented in the appendices.

%%%%%%%%%%%%%%%%%%%%%%%%%%%%%%%%%%%%%%%%%%%%%%%%%%%%%%%%%%
\section{Review of surface charges in Chern-Simons theory}
\label{sec:2}
This section is to review some well-known facts about Chern-Simons theory following the Refs~\cite{Balachandran:1991dw, Banados:1994tn}. We start from the Chern-Simons theory defined on a manifold with topology $M=\mathbb{R}\times\Sigma$, whose action is
\begin{align}
\label{CS action}
I(A)=\frac{k}{4\pi}\int_{M}\text{Tr}\left(A\wedge dA+\frac{2}{3}A\wedge A \wedge A \right).
\end{align}
In the Hamiltonian form, the action can be expressed as
\begin{align}
\label{cs-h-action}
I(A)&=\frac{k}{4\pi}\int_{\mathbb{R}} dt\int_{\Sigma}d^2x \varepsilon^{ij}g_{ab}(\dot A^a_i A^b_j+A^a_tF^b_{ij})+B,
\end{align}
where the $g_{ab}$ is the Cartan-Killing metric of the gauge group. The $B$ is a boundary term that depends on the imposed boundary condition. The boundary term plays a crucial role in the charges and the symmetry algebra~\cite{Regge:1974zd}. As a consequence, we may get the different charges and symmetries by imposing various boundary conditions in Chern-Simons theory.
\par
From the Hamiltonian form, we learn that the $A^a_i$ are the dynamics fields and $A^a_t$ is the Lagrange multiplier. Varying with respect to $A^a_i$, one can get the equation of motion
\begin{align}
F_{ti}=\partial_tA_i^a-\partial_{i}A_t^a+f^a_{bc}A_t^bA_{i}^c=0.
\end{align}
The Lagrange multiplier gives the constraint
\begin{align}\label{2.4}
G_a\equiv\frac{k}{4\pi}g_{ac}\varepsilon^{ij}F^c_{ij}=0.
\end{align}
The canonical momenta of the dynamical fields $A_i^a$ are $A_{j}^b$, which satisfy the canonical Poisson bracket
\begin{align}
\{A^a_i(x),A^b_j(y)\}=\frac{2\pi}{k}g^{ab}\varepsilon_{ij}\delta(x-y).
\end{align}
One can choose different boundary conditions, there will be different boundary terms $B$ in~\eqref{cs-h-action} but the canonical Poisson bracket does not change.
The Poisson bracket for any two functions $G,H$ can be calculated by
\begin{align}
\{G(A^a_i),H(A^b_j)\}=\frac{2\pi}{k}\int d^2x\varepsilon_{ij}g^{ab}\frac{\delta G}{\delta A^a_i}\frac{\delta H}{\delta A^b_j}.
\end{align}
Therefore, one can find the constraints satisfy the Poisson algebra
\begin{align}\label{2.7}
\{G_a(x), G_b(y)\}=f^c_{ab}G_c(x)\delta(x-y),
\end{align}
which implies $G_a=0$ are the first class constraints.
\par
Moreover, we should also consider the smeared generator
\begin{align}
\label{CS-charges}
G(\eta)=\int_{\Sigma}G_a\eta^a +Q(\eta),\quad Q(\eta)=-\frac{k}{2\pi}\int_{\partial\Sigma}\eta_aA^a.
\end{align}
The supplemented term $Q(\eta)$ is to make the smeared generator differentiable~\cite{Balachandran:1991dw}. In general, the parameter $\eta$ is the set of gauge transformations that preserve the imposed boundary conditions. The Poisson bracket of the smeared generators is
\begin{align}
\{G(\eta),G(\lambda)\}&=G([\eta,\lambda])+C(\eta,\lambda),\quad C(\eta,\lambda)=\frac{k}{2\pi}\int_{\partial\Sigma}\eta_ad\lambda^a,\nonumber
\end{align}
where $C(\eta,\lambda)$ is the central charge term. As a consequence, the Poisson bracket of the smeared generator is a central extension of the algebra of the gauge generator (\ref{2.7}). The central extension comes from the surface term $Q(\eta)$ in the definition of the smeared generator. It is worth noting that the smeared generator does not always vanish when the constraints $G_a=0$ are imposed. The transformations generated by $G(\eta)$ are not always the trivial gauge transformations, which can transform one physical state to another~\cite{Banados:1998gg}. While the trivial gauge transformations correspond to vanishing $G(\eta)$. The asymptotic symmetry, also called global symmetry, is defined as the quotient of the group of gauge transformations modulo the group of the trivial gauge transformations. This is the origin of infinitely many boundary degrees of freedom in Chern-Simons theory.
\par
After disentangling the constraints (\ref{2.4}), the $Q(\eta)$ define the surface charges of the Chern-Simons theory. It turns out the surface charges satisfy the same Poisson bracket algebra
\begin{align}
\{Q(\eta),Q(\lambda)\}=Q([\eta,\lambda])+C(\eta,\lambda).
\end{align}
Furthermore, the variation of a function in phase space can be generate by the surface charges
\begin{align}
\delta_{\lambda}F=\{Q(\lambda),F\}
\end{align}
Given a certain boundary condition, we can find the gauge transformations preserving the boundary condition. The corresponding charges induced by the boundary condition can also be obtained. This technique was widely used in AdS$_3$ with various boundary conditions~\cite{Banados:1998gg,Compere:2013bya,Compere:2015knw,Grumiller:2016pqb,Afshar:2016wfy,Afshar:2016kjj,Perez:2016vqo,Ojeda:2019xih,Sheikh-Jabbari:2016unm}. In this paper, we would like to apply this approach to study the surface charges of the Chern-Simons gravity theory with the mixed boundary condition for $T\bar{T}$ deformation~\cite{Guica:2019nzm}.

%%%%%%%%%%%%%%%%%%%%%%%%%%%%%%%%%%%%%%%%%%%%%%
\section{Chern-Simons formalism and $T\bar{T}$ deformation}
\label{sec:3}
%%%%%%%%%%%%%%%%%%%%%%%%%%%%%%%%%%%%%%%%%%%%%%%%%%%%%%%%%%%%%%%%%%
The AdS$_3$ gravity can be formulated as $SL(2,\mathbb R)\times SL(2,\mathbb R)$ Chern-Simons theory~\cite{Witten:1988hc}. The action can be written as the sum of the left-moving part and right-moving part
\begin{align}
S(A,\bar A)=I(A)-I(\bar A),\quad\text{with}\quad k=\frac{1}{4G},
\end{align}
where the gauge fields are the combination of vielbein and spin connection
\begin{align}
A^a=\omega^a+e^a,\quad\bar A^a=\omega^a-e^a.
\end{align}
The equations of motion are
\begin{align}
\label{equations of motion}
dA+A\wedge A=0,\quad d\bar A+\bar A\wedge\bar A=0.
\end{align}
which agree with first order gravitational field equations. Given an AdS$_3$ solution, we have an equivalent description in the Chern-Simons formalism.
\par
In particular, the Ba\~nados geometry~\cite{Banados:1998gg} in Fefferman-Graham gauge is following
\begin{align}
ds^2=\frac{dr^2}{r^2}+r^2\left(dzd\bar z+\frac{1}{r^2}\mathcal{L}(z)dz^2+\frac{1}{r^2}\mathcal{\bar L}(\bar z)d\bar z^2+\frac{1}{r^4}\mathcal{L}(z)\mathcal{\bar L}(\bar z)dzd\bar z\right),
\end{align}
where the $\mathcal{L}(z)$ and $\mathcal{\bar{L}}(\bar{z})$ are arbitrary holomorphic and antiholomorphic functions, respectively. For the case of BTZ black hole, the parameters are constants associated with the mass and angular momentum of the black hole
\begin{align}
\label{BTZrelation}
\mathcal{L}=\frac{M+J}{2},\quad \mathcal{\bar{L}}=\frac{M-J}{2}.
\end{align}
Up to the Lorentz rotation, the corresponding Chern-Simons gauge connections can be fixed as
\begin{align}
\tilde{A}=&\frac{dr}{r}L_0+rdzL_{-1}+\frac{1}{r}\mathcal{L}dzL_1,\\
\tilde{\bar A}=&-\frac{dr}{r}L_0-\frac{1}{r}\mathcal{\bar L}d\bar zL_{-1}-rd\bar zL_1.
\end{align}
where the $L_{-1},L_0,L_1$ are the generators of $SL(2,\mathbb{R})$. In this paper, we use the  following generators of $SL(2,\mathbb R)$
\begin{align}
L_{-1}=\left(
  \begin{matrix}
   0 &\  0 \\
   1 &\  0
  \end{matrix}
  \right),
  L_{0}=\frac{1}{2}\left(
  \begin{matrix}
   1 &\  0 \\
   0 &\  -1
  \end{matrix}
  \right),
  L_{1}=\left(
  \begin{matrix}
   0 &\  1 \\
   0 &\  0
  \end{matrix}
  \right),
\end{align}
with the commutation relations\footnote{In our convention, we use the different sign of $L_1$ rather than the usual convention in~\cite{Grumiller:2016pqb,Bunster:2014mua}. The commutation relations of the generators become~\eqref{commutation-relations}}.
\begin{align}
\label{commutation-relations}
[L_{-1},L_0]=L_{-1},\quad [L_{-1},L_1]=-2L_0,\quad [L_{0},L_1]=L_1.
\end{align}
The non-zero components of Cartan-Killing metric are
\begin{align}
\text{Tr}\left(L_{-1}L_{1}\right)=\text{Tr}\left(L_{1}L_{-1}\right)=1,\quad \text{Tr}\left(L_{0}L_{0}\right)=\frac{1}{2}.
\end{align}
In this paper, we will use the $(\tilde{A},\tilde{\bar{A}})$ denote the original gauge fields and $(A,\bar{A})$ denote the deformed gauge fields.
Following~\cite{Coussaert:1995zp}, the $r$-dependent of the gauge fields can be eliminated through a gauge transformation
\begin{align}
\label{radial_gauge}
\tilde{A}=b^{-1}(d+\tilde{a})b,\quad \tilde{\bar{A}}=b(d+\tilde{\bar{a}})b^{-1},\quad b=e^{\ln rL_0}.
\end{align}
The induced gauge fields take the form
\begin{align}
\tilde{a}=(L_{-1}+\mathcal{L}L_{1})dz,\quad \tilde{\bar{a}}=(\mathcal{\bar{L}}L_{-1}+L_{1})d\bar{z},
\end{align}
which can be treated as the gauge connection defined on the boundary. For the Ba\~nados geometry, the residual gauge symmetry generates the Virasoro algebra~\cite{Banados:1998gg}.
\par
The $T\bar{T}$ deformed CFTs correspond to the AdS$_3$ gravity with a mixed boundary condition~\cite{Guica:2019nzm,Llabres:2019jtx}. The deformed AdS$_3$ solutions can also be constructed from original one via a field-dependent coordinate transformation~\cite{Guica:2019nzm}. For the Ba\~nados geometry, the field-dependent coordinate transformation reads
\begin{align}
\label{coordinate transformation}
dz=\frac{1}{1-\mu^2\mathcal{L_{\mu}\bar L_{\mu}}}(dw-\mu\mathcal{\bar L}_{\mu}d\bar w),\quad d\bar z=\frac{1}{1-\mu^2\mathcal{L_{\mu}\bar{L}_{\mu}}}(d\bar w-\mu\mathcal{L}_{\mu}dw),
\end{align}
where $\mathcal{L}_{\mu}\equiv\mathcal{L}(z(\mu,w,\bar{w})),\mathcal{\bar{L}}_{\mu}\equiv\mathcal{L}(z(\mu,w,\bar{w}))$, and $\mu$ is the deformation parameter. Then, we can obtain the deformed the gauge fields
\begin{align}
A=&\frac{1}{r}L_0dr+\frac{1}{1-\mu^2\mathcal L_{\mu}\mathcal {\bar L}}_{\mu}\left(rL_{-1}+\frac{1}{r}\mathcal{L}_{\mu}L_1\right)(dw-\mu \mathcal{\bar L}_{\mu}d\bar{w}),\\
\bar{A}=&-\frac{1}{r}L_0dr-\frac{1}{1-\mu^2\mathcal L_{\mu}\mathcal {\bar L}_{\mu}}\left(\frac{1}{r}\mathcal{\bar L}_{\mu}L_{-1}+rL_{1}\right)(d\bar{w}-\mu \mathcal{L}_{\mu}dw).
\end{align}
The Ba\~nados geometry is parametrized by the holomorphic function $\mathcal{L}(z)$ and anti-holomorphic function $\mathcal{\bar{L}}(\bar{z})$. The deformed metric is parametrized by $\mathcal{L}_{\mu}$ and $\mathcal{\bar{L}}_{\mu}$. The coordinate transformation implies the deformed parameter $\mathcal{L}_{\mu}$ and $\mathcal{\bar{L}}_{\mu}$ obey
\begin{align}
\label{L_evolution}
\partial_{\bar{w}}\mathcal{L}_{\mu}+\mu\mathcal{\bar{L}}_{\mu}\partial_{w}\mathcal{L}_{\mu}=0,\\
\label{Lbar_evolution}
\partial_{w}\mathcal{\bar{L}}_{\mu}+\mu\mathcal{L}_{\mu}\partial_{w}\mathcal{\bar{L}}_{\mu}=0.
\end{align}
Since these gauge connections satisfy the equation of motions, the deformed metrics are still the solution of AdS$_3$.
In~\cite{Guica:2019nzm}, it is also shown that the deformed parameters are following
\begin{align}
\label{LLbar_relations}
\mathcal{L}=\frac{\mathcal{L}_{\mu}(1-\mu\mathcal{\bar{L}}_{\mu})^2}{(1-\mu^2\mathcal{L_{\mu}\bar L_{\mu}})^2},\quad \mathcal{\bar{L}}=\frac{\mathcal{\bar{L}}_{\mu}(1-\mu\mathcal{L}_{\mu})^2}{(1-\mu^2\mathcal{L_{\mu}\bar L_{\mu}})^2}.
\end{align}
\par
We prefer to use the coordinates $\theta=(w+\bar{w})/2,t=(w-\bar{w})/2$, where $t$ represents the time direction while $\theta$ represents a circle at the boundary with the identification $\theta\sim\theta+2\pi$. In this coordinate, the gauge fields can be written as
\begin{align}
A_{r}=&\frac{1}{r}L_0,\quad \bar{A}_{r}=-\frac{1}{r}L_0\\
A_{\theta}=&\frac{1-\mu\mathcal{\bar{L}}_{\mu}}{1-\mu^2\mathcal{L}_{\mu}\mathcal{\bar L}_{\mu}}(rL_{-1}+\frac{1}{r}\mathcal{L}_{\mu}L_1),\quad A_{t}=KA_{\theta},\\
\bar {A}_{\theta}=&-\frac{1-\mu \mathcal {L}_{\mu}}{1-\mu^2\mathcal {L}_{\mu}\mathcal {\bar L}_{\mu}}(\frac{1}{r}\mathcal {\bar L}_{\mu}L_{-1}+rL_1),\quad \bar{A}_{t}=\bar{K}\bar{A}_{\theta}.
\end{align}
where we define
\begin{align}
K=\frac{1+\mu\mathcal{\bar{L}}_{\mu}}{1-\mu\mathcal{\bar{L}}_{\mu}},\quad \bar{K}=-\frac{1+\mu\mathcal{L}_{\mu}}{1-\mu\mathcal{L}_{\mu}}.
\end{align}
\par
In the Ba\~nados geometry, the parameters $\mathcal{L}$ and $\mathcal{\bar{L}}$ relate to the charges. The new parameters $\mathcal{L}_{\mu}$ and $\mathcal{\bar{L}}_{\mu}$ do not play the role of charges in the deformed geometry. In analogy with the un-deformed case, we can construct the new parameters using the spectrum and angular momentum. The deformed spectrum and angular momentum can be obtained from gravity side~\cite{McGough:2016lol, Guica:2019nzm}. In Chern-Simons form, the boundary term consistent with the mixed boundary condition turns out to be
\begin{align}
B=&\frac{\kappa}{4\pi}\int_{\partial M}dtd \theta\frac{1}{\mu}\left(\sqrt{1-2\mu\left(X_{\theta\theta}+\bar X_{\theta\theta}\right)+\mu^2\left(X_{\theta\theta}-\bar X_{\theta\theta}\right)^2}-1\right).
\end{align}
where
\begin{align}
X_{ij}=\text{Tr}(A_{i}A_{j}),\quad \bar{X}_{ij}=\text{Tr}(\bar A_{i}\bar A_{j}).
\end{align}
One refer to~\cite{Llabres:2019jtx} for general form of the boundary term. The resulting deformed spectrum and angular momentum are as follows 
\begin{align}
\mathcal{E}=\frac{1}{\mu}\left(1-\sqrt{1-2\mu(\mathcal{L}+\mathcal{\bar{L}})+\mu^2(\mathcal{L}-\mathcal{\bar{L}})^2}\right),\quad \mathcal{J}=\mathcal{L}-\mathcal{\bar{L}}.
\end{align}
We then introduce the new parameters 
\begin{align}
q=\frac{\mathcal{E}+\mathcal{J}}{2},\quad \bar{q}=\frac{\mathcal{E}-\mathcal{J}}{2},
\end{align}
In section~\ref{sec:4}, the $q$ and $\bar{q}$ are turned out to be the surface charges. The charges reduced to the Virasoro charges when $\mu\to 0$. As a consequence, we have three ways to parametrize the deformed gauge fields, by $(\mathcal{L},\mathcal{\bar{L}})$, $(\mathcal{L}_{\mu},\mathcal{\bar{L}}_{\mu})$ and $(q,\bar{q})$. The relations between different parameters are
\begin{align}
\label{eta}
\frac{1-\mu\mathcal{\bar L}_{\mu}}{1-\mu^2\mathcal{L}_{\mu}\mathcal{\bar L}_{\mu}}=&\frac{1}{2}\left[1+\mu(\mathcal{L}-\mathcal{\bar{L}})+\sqrt{1-2\mu(\mathcal{L}+\mathcal{\bar{L}})+\mu^2(\mathcal{L}-\mathcal{\bar{L}})^2}\right]=1-\mu\bar{q},\\
\label{etabar}
\frac{1-\mu\mathcal{L}_{\mu}}{1-\mu^2\mathcal{L}_{\mu}\mathcal{\bar L}_{\mu}}=&\frac{1}{2}\left[1-\mu(\mathcal{L}-\mathcal{\bar{L}})+\sqrt{1-2\mu(\mathcal{L}+\mathcal{\bar{L}})+\mu^2(\mathcal{L}-\mathcal{\bar{L}})^2}\right]=1-\mu q.
\end{align}
In the latter of this paper, we will use different parameters to simplify the expressions, and they can be transformed to each other with the help of the above relations.
\par
Finally, the deformed gauge connection can be expressed in terms of $(q,\bar{q})$
\begin{align}
\label{gauge field A}
A_{\theta}=&r(1-\mu\bar q)L_{-1}+\frac{1}{r}qL_1,\quad A_{t}=K\Big(r(1-\mu\bar q)L_{-1}+\frac{1}{r} qL_1\Big),\\
\label{gauge field Abar}
\bar A_{\theta}=&-\frac{1}{r}\bar q L_{-1}-r(1-\mu q)L_1,\quad \bar A_{t}=-\bar{K}\Big(\frac{1}{r}\bar q L_{-1}+r(1-\mu q)L_1\Big),
\end{align}
with
\begin{align}
K=&\frac{1+\mu(\bar q-q)}{1-\mu(q+\bar q)},\quad \bar{K}=-\frac{1-\mu(\bar q-q)}{1-\mu(q+\bar q)}.
\end{align}
Moreover,~\eqref{L_evolution} and~\eqref{Lbar_evolution} imply the parameters $(q,\bar{q})$ satisfy the equations
\begin{align}
\label{q-evolution}
\partial_{t}q=&\partial_{\theta}(Kq),\\
\label{qbar-evolution}
\partial_{t}\bar{q}=&\partial_{\theta}(\bar{K}\bar{q}),
\end{align}
from which we can also see that the deformed charges are no longer holomorphic or antiholomorphic. When taking the limit $\mu\to 0$, the gauge connection would reduce to the undeformed case. For the deformed geometry, the radial degree of freedom can also be eliminated through the gauge transformation~\eqref{radial_gauge}, the induced gauge connections are
\begin{align}
\label{gauge field a}
a_{\theta}=&(1-\mu\bar q)L_{-1}+qL_1,\quad a_{t}=K\Big((1-\mu\bar q)L_{-1}+qL_1\Big),\\
\label{gauge field abar}
\bar a_{\theta}=&-\bar q L_{-1}-(1-\mu q)L_1,\quad \bar a_{t}=-\bar{K}\Big(\bar q L_{-1}+(1-\mu q)L_1\Big).
\end{align}
These gauge fields are defined on the boundary. We then will apply the induced gauge fields to study the symmetry of $T\bar{T}$ deformation in the next section.
%%%%%%%%%%%%%%%%%%%%%%%%%%%%%%%%%%%%%%%%%%%%%%
\section{Surface charges and their algebra}
\label{sec:4}
In this section, we would like to calculate the surface charges induced by asymptotic symmetries of Chern-Simons theory with $T\bar{T}$ deformation. Firstly, we have to find the residual gauge symmetry generators of the $T\bar{T}$ deformed gauge fields. We then calculate the surface charges associated with the gauge symmetries. Finally, we obtain the algebra of the deformed surface charges.
\subsection{Boundary condition and symmetries of the deformed gauge fields}
For the deformed gauge fields, we assume the variation of the charges $q,\bar{q}$ induced by gauge transformation of the deformed gauge field as following forms
\begin{align}
\lambda:\quad q\to q+\delta_{\lambda}q,\quad \bar{q}\to\bar{q}+\delta_{\lambda}\bar{q},\\
\bar{\lambda}:\quad q\to q+\delta_{\bar{\lambda}}q,\quad \bar{q}\to\bar{q}+\delta_{\bar{\lambda}}\bar{q}.
\end{align}
Where $\lambda$ and $\bar{\lambda}$ are defined as the left-moving part and the right-moving part respectively
 \begin{align}
\lambda=\sum_{i=-1}^{1}\lambda_{i}L_{i},\quad
\bar{\lambda}=\sum_{i=-1}^{1}\bar{\lambda}_{i}L_{i}.
\end{align}
Then the variation of the gauge fields can be expressed as
\begin{align}
\delta_{\lambda} a_{\theta}=&-\mu\delta_{\lambda}\bar{q}L_{-1}+\delta _{\lambda}qL_1,\\
\delta_{\lambda} a_{t}=&\delta_{\lambda}\left(K(1-\mu\bar{q})\right)L_{-1}+\delta_{\lambda}(Kq)L_1,\\
\delta_{\bar{\lambda}}\bar{a}_{\theta}=&-\delta_{\bar{\lambda}}\bar{q}L_{-1}+\mu\delta _{\bar{\lambda}}qL_1,\\
\delta_{\bar{\lambda}}\bar{a}_{t}=&-\delta_{\bar{\lambda}}\left(\bar{K}\mu\bar{q}\right)L_{-1}-\delta_{\bar{\lambda}}(\bar{K}(1-\mu q))L_1,
\end{align}
where the variation of $K,\bar{K}$ can also expressed in terms of the variation of $q,\bar{q}$
\begin{align}
\delta_{\lambda}K=&\frac{2 \mu  \left(\mu  \bar{q} \delta_{\lambda}q+(1-\mu  q)\delta_{\lambda}\bar{q}\right)}{1-\left(\mu\left(\bar{q}+q\right)\right)^2},\\
\delta_{\bar{\lambda}}\bar{K}=&-\frac{2 \mu  \left(\left(1-\mu  \bar{q}\right)\delta_{\bar{\lambda}}q+\mu  q\delta_{\bar{\lambda}}\bar{q}\right)}{\left(1-\mu\left(\bar{q}+q\right)\right)^2}.
\end{align}
Then, we have to find the relations between $(\delta_{\lambda}q,\delta_{\lambda}\bar{q})$ and $(\delta_{\bar\lambda}q,\delta_{\bar\lambda}\bar{q})$ by using the mixed boundary condition. In our setting, the gauge fields can reproduce the deformed metric through
\begin{align}
g_{\mu\nu}=&\frac{1}{2}\text{Tr}[(A_{\mu}-\bar{A}_{\mu})(A_{\nu}-\bar{A}_{\nu})].
\end{align}
The induced boundary metric at the finite cutoff surface $r=r_c$ turns out to be  a flat one
\begin{align}
ds^2\Big|_{r=r_c}=\frac{1}{\mu}(d\theta^2-dt^2),
\end{align}
where we have invoked the holographic relation $r_c^2=1/\mu$ in $T\bar{T}$ deformation~\cite{McGough:2016lol}. It follows that the variation of the metric on the boundary should be vanishing
\begin{align}
\label{Dirichlet boundary condition}
\delta g_{\mu\nu}\Big|_{r=r_c}=\text{Tr}[(\delta_{\lambda} A_{\mu}-\delta_{\bar{\lambda}}\bar{A}_{\mu})(A_{\nu}-\bar{A}_{\nu})]\Big|_{r=r_c}=0.
\end{align}
{It means that the residual gauge symmetries become the exact  symmetries on the surface $r=r_c$.} The equation~\eqref{Dirichlet boundary condition} gives
\begin{align}
&(\delta_{\bar{\lambda}}\bar{q}-\delta_{\lambda}\bar{q})+(\delta_{\lambda} q-\delta_{\bar{\lambda}}q)=0,\\
&(\delta_{\bar{\lambda}}\bar{q}-\delta_{\lambda}\bar{q})-(\delta_{\lambda} q-\delta_{\bar{\lambda}}q)=0,\\
&\delta_{\lambda}(K(1-\mu\bar{q}))+\delta_{\bar\lambda}(\mu K\bar{q})+\delta_{\lambda}(\mu Kq)+\delta_{\bar{\lambda}}(\bar{K}(1-\mu q))=0,\\
&\delta_{\lambda}(K(1-\mu\bar{q}))+\delta_{\bar\lambda}(\mu K\bar{q})-\delta_{\lambda}(\mu Kq)-\delta_{\bar{\lambda}}(\bar{K}(1-\mu q))=0.
\end{align}
The unique solution for these equations implies the constraints
\begin{align}
\label{variation_relation_1}
\delta_{\lambda} q-\delta_{\bar{\lambda}}q=&0,\\
\label{variation_relation_2}
\delta_{\bar{\lambda}}\bar{q}-\delta_{\lambda}\bar{q}=&0.
\end{align}
These relations mean that the gauge transformations on the left-moving part and right-moving part are entangled.
\par
Under the infinitesimal gauge transformation, variation of the deformed gauge fields are
\begin{align}
\delta_{\lambda}a=d\lambda+[a,\lambda],\quad \delta_{\bar{\lambda}}\bar{a}=d\bar{\lambda}+[\bar{a},\bar{\lambda}].
\end{align}
The gauge transformation that preserve the asymptotic behavior of $a_{\theta},\bar{a}_{\theta}$ gives
\begin{align}
\label{meq-lambda-1}
-\mu\delta_{\lambda}\bar q=&\lambda'_{-1}+\lambda_0(1-\mu \bar q),\\
\label{meq-lambda0}
0=&\lambda'_0+2\Big(\lambda_{-1}q-\lambda_1(1-\mu \bar q)\Big),\\
\label{meq-lambda1}
\delta_{\lambda} q=&\lambda'_1-\lambda_0q,\\
\label{meq-lambdabar-1}
-\delta_{\bar{\lambda}}\bar q=&\bar\lambda_{-1}'-\bar\lambda_0\bar q,\\
\label{meq-lambdabar0}
0=&\lambda'_0+2\Big(\bar\lambda_1\bar q-\bar\lambda_{-1}(1-\mu q)\Big),\\
\label{meq-lambdabar1}
\mu\delta_{\bar{\lambda}} q=&\bar\lambda'_1+\bar\lambda_0(1-\mu q).
\end{align}
The gauge transformation that preserves the asymptotic behavior of $a_{t},\bar{a}_{t}$ gives
\begin{align}
\label{mcons_lambda-1}
\delta_{\lambda}\left(K(1-\mu\bar{q})\right)=&\partial_{t}\lambda_{-1}+K\lambda_{0}(1-\mu\bar{q}),\\
\label{mcons_lambda0}
0=&\partial_{t}\lambda_{0}+2K\Big(\lambda_{-1}q-\lambda_{1}(1-\mu\bar{q})\Big),\\
\label{mcons_lambda1}
\delta_{\lambda}\left(Kq\right)=&\partial_{t}\lambda_{1}-K\lambda_{0}q,\\
\label{mcons_lambdabar-1}
-\delta_{\bar{\lambda}}(\bar{K}\bar{q})=&\partial_{t}\bar{\lambda}_{-1}-\bar{K}\bar{\lambda}_{0}\bar{q},\\
\label{mcons_lambdabar0}
0=&\partial_{t}\bar{\lambda}_{0}+2\bar{K}\Big(\bar{\lambda}_{1}\bar{q}-\bar{\lambda}_{-1}(1-\mu q)\Big),\\
\label{mcons_lambdabar1}
-\delta_{\bar{\lambda}}\left(\bar{K}(1-\mu q)\right)=&\partial_{t}\bar{\lambda}_{1}+\bar{K}\bar{\lambda}_{0}(1-\mu q).
\end{align}
\par
For later convenience, one can choose the following parameters
\begin{align}
\epsilon=\lambda_{-1}-\mu\bar{\lambda}_{-1},\quad \bar \epsilon=\bar\lambda_{1}-\mu\lambda_{1},
\end{align}
to parametrize the gauge transformation generators. By solving the equations~\eqref{meq-lambda-1}-\eqref{meq-lambdabar1}, we can express the variation of the charges $\delta_{\lambda}q$ and $\delta_{\bar{\lambda}}\bar{q}$ in terms of the parameters $(\epsilon,\bar{\epsilon})$
\begin{align}
\label{variation q}
\delta_{\lambda}q=\delta_{\bar\lambda}q=&\epsilon(\theta)C'+\bar{\epsilon}(\theta)A'+\frac{1}{2}\bar{\epsilon}'(\theta)B''-\frac{1}{2}\epsilon'(\theta)(B''-4C)\nonumber\\
&-\frac{1}{2}\bar{\epsilon}''(\theta)(D'-B')-\frac{3}{2}\epsilon''(\theta)B'-\frac{1}{2}\bar{\epsilon}'''(\theta )D-\frac{1}{2}\epsilon'''(\theta )(2B+1),\\
\label{variation qbar}
\delta_{\lambda}\bar{q}=\delta_{\bar{\lambda}}\bar{q}=&-\epsilon(\theta)\bar{A}'-\bar{\epsilon}(\theta)\bar{C}'-\frac{1}{2}\epsilon'(\theta)\bar{B}''+\frac{1}{2}\bar{\epsilon}'(\theta)(\bar{B}''-4\bar{C})\nonumber\\
&+\frac{1}{2}\epsilon''(\theta)(\bar{D}'-\bar{B}')+\frac{3}{2}\bar{\epsilon}''(\theta)\bar{B}'+\frac{1}{2}\epsilon'''(\theta )\bar{D}+\frac{1}{2}\bar{\epsilon}'''(\theta )(2\bar{B}+1),
\end{align}
where we have introduced the following auxiliary variables to simplify the expressions
\begin{align}
A=&\bar{A}=\frac{\mu q \bar{q}}{1-\mu\left(\bar{q}+q\right)},\\
B=&\frac{\mu\left(2\bar{q}-\mu(q+\bar q)^2\right)}{2(1-\mu(q+\bar q))^2},\quad \bar B=\frac{\mu\left(2q-\mu(q+\bar q)^2\right)}{2(1-\mu(q+\bar q))^2},\\
C=&\frac{(1-\mu q)q}{1-\mu(q+\bar q)},\quad \bar{C}=\frac{(1-\mu \bar q)\bar q}{1-\mu(q+\bar q)},\\
D=&-\bar{D}=\frac{\mu  \left(q-\bar{q}\right)}{\left(1-\mu\left(\bar{q}+q\right)\right)^2}.
\end{align}
One can turn to the Appendix~\ref{app-A} for details about solving the variation of the charges.
\par
From~\eqref{mcons_lambda-1}-\eqref{mcons_lambdabar1}, one can find the variables $\epsilon$ and $\bar{\epsilon}$ satisfy
\begin{align}
\label{epsion}
-\partial_{t}\epsilon=&\epsilon'-\frac{2 \mu  \bar{q} \left(1-\mu  \bar{q}\right)}{\left(1-\mu \left(\bar{q}+q\right)\right)^2}
\left(\epsilon'+\bar{\epsilon}'\right),\\
\label{epsionbar}
\partial_{t}\bar{\epsilon}=&\bar{\epsilon}'-\frac{2 \mu q\left(1-\mu  q\right)}{\left(1-\mu  \left(\bar{q}+q\right)\right)^2}\left(\epsilon'+\bar{\epsilon}'\right).
\end{align}
The details for deriving these equations are given in Appendix~\ref{app-B}. In~\cite{Kraus:2021cwf}, the same result was obtained by using the Killing vector that leaves all components of the metric invariant at the cutoff boundary. For the undeformed case, namely $\mu\to0$, the equations reduce to that the $\epsilon$ and $\bar{\epsilon}$ are holomorphic and antiholomorphic functions, respectively, which indeed correspond to the infinitesimal conformal transformation. The equations~\eqref{epsion} and~\eqref{epsionbar} can be identified as the $T\bar{T}$ deformed conformal Killing equations.

\subsection{Surface charges and their algebra}
Given the gauge transformation preserving the asymptotic behavior of gauge fields, we can obtain the associated surface charge using~\eqref{CS-charges}. The variation of the charges spanned
by gauge parameters $\lambda$ and $\bar{\lambda}$ in the Chern-Simons formalism read
\begin{align}
\delta \mathcal{Q}_{\epsilon,\bar{\epsilon}}=&\int_{\partial\Sigma}\left(\text{Tr}(\lambda\delta A_{\theta})-\text{Tr}(\bar{\lambda}\delta \bar{A}_{\theta})\right)d\theta\nonumber\\
=&\int_{\partial\Sigma}\left(-\mu\lambda_{1}(\delta_{\lambda}\bar{q}+\delta_{\bar{\lambda}}\bar{q})+\lambda_{-1}(\delta_{\lambda}q+\delta_{\bar{\lambda}}q)\right)d\theta\nonumber\\
&-\int_{\partial\Sigma}\left(-\bar{\lambda}_{1}(\delta_{\lambda}\bar{q}+\delta_{\bar{\lambda}}\bar{q})+\mu\bar\lambda_{-1}(\delta_{\lambda}q+\delta_{\bar{\lambda}}q)\right)d\theta\nonumber\\
=&\int_{\partial\Sigma}2\left(\delta_{\lambda} q\epsilon-\delta_{\bar\lambda}\bar{q}\bar{\epsilon}\right)d\theta.
\end{align}
The charges $\mathcal{Q}_{\epsilon,\bar{\epsilon}}$ are the generator of residual symmetries which combined $\lambda$ and $\bar{\lambda}$. We divide the variations of $q$ and $\bar{q}$ into two parts in the second step. In the third step, we have used equations~\eqref{variation_relation_1} and~\eqref{variation_relation_2}. The charges can be defined as
\begin{align}
\mathcal{Q}_{\epsilon,\bar{\epsilon}}
=&\int_{0}^{2\pi}2\Big(q(\theta)\epsilon(\theta)-\bar{q}(\theta)\bar{\epsilon}(\theta)\Big)d\theta.
\end{align}
With help of the parameter $\epsilon$ and $\bar{\epsilon}$, the charges split into two independent parts. For convenience, we can define the charges
\begin{align}
Q=\frac{1}{2}\mathcal{Q}_{\epsilon}=\int_{0}^{2\pi}q(\theta)\epsilon(\theta)d\theta,\\
\bar{Q}=\frac{1}{2}\mathcal{Q}_{\bar{\epsilon}}=\int_{0}^{2\pi}\bar{q}(\theta)\bar{\epsilon}(\theta)d\theta.
\end{align}
\par
The variation of the charges under the symmetry transformation can be expressed as Poisson bracket algebra
\begin{align}
\delta_{\lambda}q=\delta_{\bar{\lambda}}q=\frac{1}{2}\delta_{\lambda,\bar{\lambda}}q=\frac{1}{2}\{\mathcal{Q}_{\epsilon,\bar{\epsilon}},q\}=\int_{0}^{2\pi}\Big(\{q(\theta'),q(\theta)\}\epsilon(\theta')-\{\bar{q}(\theta'),q(\theta)\}\bar{\epsilon}(\theta')\Big)d\theta',\\
\delta_{\lambda}\bar{q}=\delta_{\bar{\lambda}}\bar{q}=\frac{1}{2}\delta_{\lambda,\bar{\lambda}}\bar{q}=\frac{1}{2}\{\mathcal{Q}_{\epsilon,\bar{\epsilon}},\bar{q}\}=\int_{0}^{2\pi}\Big(\{q(\theta'),\bar{q}(\theta)\}\epsilon(\theta')-\{\bar{q}(\theta'),\bar{q}(\theta)\}\bar{\epsilon}(\theta')\Big)d\theta',
\end{align}
which allows us to identify the Poisson brackets of $q$ and $\bar{q}$ on the phase space of asymptotically AdS$_3$ solutions. According to~\eqref{variation q} and~\eqref{variation qbar}, after performing integration by parts and dropping some total derivative terms, we obtain the Poisson brackets
\begin{align}
i\{q(\theta'),q(\theta)\}=&C'\delta(\theta-\theta')-\frac{1}{2}(B''-4C)\delta'(\theta-\theta')-\frac{3}{2}B'\delta''(\theta-\theta')  -\frac{1}{2}(2B+1)\delta'''(\theta-\theta'),\\
i\{\bar{q}(\theta'),q(\theta)\}=&-A'\delta(\theta-\theta')-\frac{1}{2}B''\delta'(\theta-\theta')+\frac{1}{2}(D'-B')\delta''(\theta-\theta')+\frac{1}{2}D\delta'''(\theta-\theta'),\\
i\{q(\theta'),\bar{q}(\theta)\}=&-\bar{A}'\delta(\theta-\theta')-\frac{1}{2}\bar{B}''\delta'(\theta-\theta')+\frac{1}{2}(\bar{D}'-\bar{B}')\delta''(\theta-\theta')+\frac{1}{2}\bar{D}\delta'''(\theta-\theta'),\\
i\{\bar{q}(\theta'),\bar{q}(\theta)\}=&\bar{C}'\delta(\theta-\theta')-\frac{1}{2}(\bar{B}''-4\bar{C})\delta'(\theta-\theta')-\frac{3}{2}\bar{B}'\delta''(\theta-\theta')-\frac{1}{2}(2\bar{B}+1)\delta'''(\theta-\theta').
\end{align}
The modes expansion of the charges is following
\begin{align}
\label{Fourier modes}
Q_{n}=\int_{0}^{2\pi}q(\theta)e^{-in\theta}d\theta,\quad \bar{Q}_{m}=\int_{0}^{2\pi}\bar{q}(\theta)e^{-im\theta}d\theta.
\end{align}
We arrive at the following Poisson brackets
\begin{align}
i\{Q_{n},Q_{m}\}=&(n-m)C_{n+m}-\frac{1}{2}mn(n-m)B_{n+m}+\frac{1}{2}n^3\delta_{n+m,0},\\
i\{Q_{n},\bar{Q}_{m}\}=&(n+m)A_{n+m}-\frac{1}{2}mn(n+m)B_{n+m}+\frac{1}{2}m^2nD_{n+m}\nonumber\\
=&(n+m)\bar{A}_{n+m}-\frac{1}{2}mn(n+m)\bar{B}_{n+m}+\frac{1}{2}mn^2\bar{D}_{n+m},\\
i\{\bar{Q}_{n},\bar{Q}_{m}\}=&(n-m)\bar{C}_{n+m}-\frac{1}{2}mn(n-m)\bar{B}_{n+m}+\frac{1}{2}n^3\delta_{n+m,0},
\end{align}
where
\begin{align}
A_{n+m}=&\bar{A}_{n+m}=\int_{0}^{2\pi}A(\theta)e^{-i(m+n)\theta}d\theta=\int_{0}^{2\pi}\frac{\mu q\bar{q}}{1-\mu(q+\bar q)}e^{-i(m+n)\theta}d\theta,\\
B_{n+m}=&\int_{0}^{2\pi}B(\theta)e^{-i(m+n)\theta}d\theta=\frac{\mu}{2}\int_{0}^{2\pi}\frac{2\bar{q}-\mu(q+\bar q)^2}{(1-\mu(q+\bar{q}))^2}e^{-i(m+n)\theta}d\theta,\\
C_{n+m}=&\int_{0}^{2\pi}C(\theta)e^{-i(m+n)\theta}d\theta=\int_{0}^{2\pi}\frac{(1-\mu q)q}{1-\mu(q+\bar q)}e^{-i(m+n)\theta}d\theta,\\
\bar{B}_{n+m}=&\int_{0}^{2\pi}\bar{B}(\theta)e^{-i(m+n)\theta}d\theta=\frac{\mu}{2}\int_{0}^{2\pi}\frac{2q-\mu(q+\bar q)^2}{(1-\mu(q+\bar{q}))^2}e^{-i(m+n)\theta}d\theta,\\
\bar{C}_{n+m}=&\int_{0}^{2\pi}\bar{C}(\theta)e^{-i(m+n)\theta}d\theta=\int_{0}^{2\pi}\frac{(1-\mu\bar{q})\bar{q}}{1-\mu(q+\bar q)}e^{-i(m+n)\theta}d\theta,\\
D_{n+m}=&-\bar{D}_{n+m}=\int_{0}^{2\pi}D(\theta)e^{-i(m+n)\theta}d\theta=\int_{0}^{2\pi}\frac{\mu(q-\bar{q})}{(1-\mu(q+\bar q))^2}e^{-i(m+n)\theta}d\theta.
\end{align}
This algebra coincides with the result in~\cite{Kraus:2021cwf}. In~\cite{Kraus:2021cwf}, the authors consider the 3D gravity in a box with Dirichlet boundary conditions. The boundary charges associated with the boundary preserving vectors give the deformed Virasoro algebra. We obtain the same result from Chern-Simons gravity with mixed boundary conditions. The quantization of this algebra is also studied in~\cite{Kraus:2021cwf,Datta:2021kha}. Taking the limit of $\mu\to 0$, this algebra reduces to the Virasoro algebra. For the non-zero $\mu$, the deformed algebra turns on a deformation of the Virasoro algebra. The central charge can be restored by multiplying the third order of $(m,n)$ by the Chern-Simons level $k=c/6$.
\par
An analogy to the Virasoro algebra, we find the zero modes of the charges give the Hamiltonian and momentum
\begin{align}
H=Q_0+\bar{Q}_0,\quad P=Q_0-\bar{Q}_0.
\end{align}
From~\eqref{q-evolution} and~\eqref{qbar-evolution}, one can obtain
\begin{align}
i\{H,Q_n\}=&\partial_{t}Q_{n},\quad i\{H,\bar{Q}_{m}\}=\partial_{t}\bar{Q}_{m},\\
i\{P,Q_n\}=&-nQ_n,\quad i\{P,\bar{Q}_m\}=-m\bar{Q}_m.
\end{align}
\par
In order to see the effect of the $T\bar{T}$ deformation, we consider the perturbative expansion of this algebra for small $\mu$. After restoring the central charge, we can obtain the first order expansion
\begin{align}
i\{Q_{n},Q_{m}\}=&(n-m)Q_{n+m}+\frac{c}{12}n^3\delta_{n+m,0}\\
&+\mu\left((n-m)(Q\bar{Q})_{n+m}-\frac{c}{12}mn(n-m)Q_{m+n}\right)+O(\mu^2),\\
i\{Q_{n},\bar{Q}_{m}\}=&\mu\left((n+m)(Q\bar{Q})_{n+m}-\frac{c}{12}(mn^2Q_{n+m}+nm^2\bar{Q}_{n+m})\right)+O(\mu^2),\\
i\{\bar{Q}_{n},\bar{Q}_{m}\}=&(n-m)\bar{Q}_{n+m}+\frac{c}{12}n^3\delta_{n+m,0}\\
&+\mu\left((n-m)(Q\bar{Q})_{n+m}-\frac{c}{12}mn(n-m)\bar{Q}_{m+n}\right)+O(\mu^2),
\end{align}
where
\begin{align}
(Q\bar{Q})_{n+m}=\sum_{k\in \mathbb{Z}}Q_{k}\bar{Q}_{n+m-k}.
\end{align}
The leading order reproduces the Virasoro algebra. The first-order correction provides a coupling between $Q_n$ and $\bar{Q}_m$. 

To close this section, we would like to point out that these charges are not conserved except for the energy and momentum\footnote{We would like to thank Per Kraus for his comments on this point.}. Even for Virasoro only $L_0$ is conserved.  For Virasoro the general $L_n$ have simple time dependence, and these charges therefore impose symmetry relations on correlators. In principle, the algebra in the cutoff case also imposes symmetry or algebra relations, but these are harder to work with due to the nonlinear properties of the algebra.  It remains to be seen whether the deformed algebra is "useful" or not.

\section{Comments on the charges}
\label{sec:5}
The charges or the symmetry generators $Q_n$ and $\bar{Q}_m$ do not remain constant with time evolutes. In other words, the non-zero modes of the charges are not conserved. As explained in~\cite{Bunster:2014mua}, the charges we obtained are not invariant but covariant under this algebra. One can refine charges by adding an explicitly time-dependent factor, such that the refined charges become time independent ones. We would like to find the explicitly time-dependent factor $X, \tilde{X}$ in equation \eqref{charge ansatz}. The strategy is to find a time-dependent modes expansion instead of~\eqref{Fourier modes} so that the refined charges still correspond to the asymptotic symmetry of AdS$_3$ with mixed boundary conditions.
\par
Following the strategy in~\cite{Bunster:2014mua}, we assume the refined charges take the following general form
\begin{align}
\label{charge ansatz}
\tilde{Q}_{n}=\int_{0}^{2\pi}qe^{-in(\theta+X)}d\theta,\quad \tilde{\bar{Q}}_{m}=\int_{0}^{2\pi}\bar{q}e^{-im(\theta+\bar{X})}d\theta
\end{align}
where $X$ and $\bar{X}$ depend on $(\theta,t)$. Then the time derivative of the charges become
\begin{align}
\partial_{t}\tilde{Q}_{n}=&\int_{0}^{2\pi}\left(\partial_{\theta}(Kq)e^{in(\theta+X)}+inqe^{in(\theta+X)}\partial_{t}X\right)d\theta,\\
\partial_{t}\tilde{\bar{Q}}_{m}=&\int_{0}^{2\pi}\left(\partial_{\theta}(\bar{K}\bar{q})e^{im(\theta+\bar{X})}+im\bar{q}e^{im(\theta+\bar{X})}\partial_{t}\bar{X}\right)d\theta,
\end{align}
which would be vanishing if we set the integrand to be a total derivative on the right hand side. One of the simple settings is
\begin{align}
\partial_{\theta}(Kq)e^{in(\theta+X)}+inqe^{in(\theta+X)}\partial_{t}X=&\partial_{\theta}(Kqe^{in(\theta+X)}),\\
\partial_{\theta}(\bar{K}\bar{q})e^{im(\theta+\bar{X})}+imqe^{im(\theta+\bar{X})}\partial_{t}\bar{X}=&\partial_{\theta}(\bar{K}\bar{q}e^{im(\theta+\bar{X})}).
\end{align}
We then obtain the equations for $X,\bar{X}$
\begin{align}
\partial_{t}X-K\partial_{\theta}X=K,\quad \partial_{t}\bar{X}-\bar{K}\partial_{\theta}\bar{X}=\bar{K}.
\end{align}
According to the coordinate transformation~\eqref{coordinate transformation}, we can write the differential operator on the left hand side as
\begin{align}
\partial_{t}-K\partial_{\theta}=-\frac{2}{1-\mu\mathcal{\bar{L}}(\bar{z})}\partial_{\bar{z}},\quad \partial_{t}-\bar{K}\partial_{\theta}=\frac{2}{1-\mu\mathcal{L}(z)}\partial_{z}.
\end{align}
where the $(z,\bar{z})$ are the original coordinates in the Ba\~nados geometry. In these coordinates, we can write down the general solutions

\begin{align}
X=&-\frac{\bar{z}}{2}-\frac{\mu}{2}\left(\int^{\bar{z}}_{0}\mathcal{\bar{L}}(\bar{z})d\bar{z}+W_{\mathcal{\bar{L}}}\right)-f(z),\quad W_{\mathcal{\bar{L}}}=N\int_{0}^{2\pi/\bar{\kappa}}\mathcal{\bar{L}}(\bar{z})d\bar{z},\\
\bar{X}=&-\frac{z}{2}-\frac{\mu}{2}\left(\int^{z}_{0}\mathcal{L}(z)dz+W_{\mathcal{L}}\right)-\bar{f}(\bar{z}),\quad W_{\mathcal{L}}=M\int_{0}^{2\pi/\kappa}\mathcal{L}(z)dz,
\end{align}
where the $f(z)$ and $\bar{f}(\bar{z})$ are arbitrary functions of $z$ and $\bar{z}$ respectively, $W_{\mathcal{\bar{L}}}$ and $W_{\mathcal{L}}$ are the  winding terms with some integers $M,N$. If the boundary is a plane, there is no winding term. If the boundary is a cylinder, the winding terms $W_{\mathcal{L}}$ and $W_{\mathcal{\bar{L}}}$ appear, and $z$ (or $\bar{z}$) ranges from $0$ to $2\pi/\kappa$ (or $2\pi/\bar{\kappa}$). Substituting $X$ and $\bar{X}$ back into~\eqref{charge ansatz}, we obtain the refined charges
\begin{align}
\label{5.10}
\tilde{Q}_{n}=&\int_{0}^{2\pi}qe^{-in\left(\theta-\frac{\bar{z}}{2}-\frac{\mu}{2}\left(\int^{\bar{z}}_{0}\mathcal{\bar{L}}(\bar{z})d\bar{z}+W_{\mathcal{\bar{L}}}\right)-f(z)\right)}d\theta,\\
\label{5.11}
\tilde{\bar{Q}}_{m}=&\int_{0}^{2\pi}\bar{q}e^{-im\left(\theta-\frac{z}{2}-\frac{\mu}{2}\left(\int^{z}_{0}\mathcal{L}(z)dz+W_{\mathcal{L}}\right)-\bar{f}(\bar{z})\right)}d\theta.
\end{align}
The result shows the refined charges are just another different kind of modes expansion. Here we have to emphasize that the two different kinds of resulting algebras
correspond to different commutation relations. One is time-dependent algebra and the other is time-independent algebra. The main difference comes from different kinds of Fourier expansion, which is closely related to state-dependence~\cite{Guica:2020uhm}. 
It seems the time-dependence of the generators~\eqref{5.10} and~\eqref{5.11} are in contradiction with the fixed time-dependence of Killing vectors, which were shown in the equations~\eqref{B12} and~\eqref{B13}. The Killing vectors have the fixed time-dependence, as argued by~\cite{Kraus:2021cwf}, one can use any initial data to define the charges on a time slice, such as equation~\eqref{Fourier modes}.  As for the refined charges, we choose different initial data for different time slices, namely time-dependent initial data. The time-dependence of the generators come from the initial date rather than the equations~\eqref{B12} and~\eqref{B13}.
\par
It is convenient to express these refined charges in terms of the original coordinates $(z,\bar{z})$. From the coordinate transformation~\eqref{coordinate transformation}, we get
\begin{align}
\theta=\frac{(w+\bar{w})}{2}=\frac{z+\bar{z}}{2}+\frac{\mu}{2}\left(\int^{\bar{z}}_{0}\mathcal{\bar{L}}(\bar{z})d\bar{z}+W_{\mathcal{\bar{L}}}+\int^{z}_{0}\mathcal{L}(z)dz+W_{\mathcal{L}}\right).
\end{align}
On a constant time slice, we find the relations
\begin{align}
 d\theta=\frac{1-\mu^2\mathcal{L}(z)\mathcal{\bar L}(\bar{z})}{1-\mu\mathcal{L}(z)}dz,\quad d\theta=\frac{1-\mu^2\mathcal{L}(z)\mathcal{\bar{L}}(\bar{z})}{1-\mu\mathcal{\bar{L}}(\bar{z})}d\bar{z}.
\end{align}
The periodicity of $\theta$ would lead to the period of $z$ and $\bar{z}$
\begin{align}
z\sim z+\frac{2\pi}{\kappa},\quad \bar{z}\sim \bar{z}+\frac{2\pi}{\bar{\kappa}},
\end{align}
where $\kappa,\bar{\kappa}$ are constants and they depend on the explicit form of $\mathcal{L}$ and $\mathcal{\bar{L}}$. In general, we can not give the specific formula of $\kappa,\bar{\kappa}$. If $\mathcal{L}$ and $\mathcal{\bar{L}}$ are constants, namely the BTZ black holes, we have
\begin{align}
\label{central charge}
\kappa=\frac{4-\mu^2(M^2-J^2)}{4-2\mu(M-J)},\quad \bar{\kappa}=\frac{4-\mu^2(M^2-J^2)}{4-2\mu(M+J)}.
\end{align}
Once the $\mu$ vanishes, the $\kappa$ and $\bar{\kappa}$ go back to the undeformed period.
\par
Finally, the refined charges end up with
\begin{align}
\label{ModchargesQ}
\tilde{Q}_{n}=&\int_{0}^{\frac{2\pi}{\kappa}}\mathcal{L}(z)e^{-in\left(\frac{z}{2}+\frac{\mu}{2}\left(\int^{z}_{0}\mathcal{L}(z)dz+W_{\mathcal{L}}\right)+f(z)\right)}dz,\\
\label{ModchargesQbar}
\tilde{\bar{Q}}_{m}=&\int_{0}^{\frac{2\pi}{\bar{\kappa}}}\mathcal{\bar{L}}(\bar{z})e^{-im\left(\frac{\bar{z}}{2}+\frac{\mu}{2}\left(\int^{\bar{z}}_{0}\mathcal{\bar{L}}(\bar{z})d\bar{z}+W_{\mathcal{\bar{L}}}\right)+\bar{f}(\bar{z})\right)}d\bar{z}.
\end{align}
In particular, one can always choose
\begin{align}
f(z)=&(\kappa-\frac{1}{2})z-\frac{\mu}{2}\left(\int^{z}_{0}\mathcal{L}(z)dz+W_{\mathcal{L}}\right),\\
\bar{f}(\bar{z})=&(\bar{\kappa}-\frac{1}{2})\bar{z}-\frac{\mu}{2}\left(\int^{\bar{z}}_{0}\mathcal{\bar{L}}(\bar{z})d\bar{z}+W_{\mathcal{\bar{L}}}\right),
\end{align}
so that the charges can be expressed as
\begin{align}
\tilde{Q}_{n}=&\int_{0}^{\frac{2\pi}{\kappa}}\mathcal{L}(z)e^{-in\kappa z}dz,\\
\tilde{\bar{Q}}_{m}=&\int_{0}^{\frac{2\pi}{\bar{\kappa}}}\mathcal{\bar{L}}(\bar{z})e^{-im\bar{\kappa}\bar{z}}d\bar{z}.
\end{align}
The refined charges are the same as the Virasoro ones, and their algebra becomes a field-dependent Virasoro algebra. This result coincides with the conclusion in~\cite{Guica:2019nzm}.
\par
In addition, for various $f(z)$, $\bar{f}(\bar{z})$ and winding terms, the refined surface charges and their algebra structure are also different. Although one can choose specific $f(z)$ and $\bar{f}(\bar{z})$ to cancel the winding terms, the charge algebra results in a certain ambiguity because of the winding terms. A similar situation happens in the field theory calculation shown in \cite{Guica:2020uhm}. It will be an interesting future problem to connect this ambiguity to the one shown in the field theory side \cite{Guica:2020uhm}.

\section{Conclusion and discussion}
\label{sec:6}
It is proposed that the $T\bar{T}$ deformed 2D CFTs dual to the cutoff AdS$_3$ with Dirichlet boundary condition or equivalently a mixed boundary condition. The mixed boundary condition can be realized by a field-dependent coordinate transformation from the Brown-Henneaux boundary condition \cite{Guica:2019nzm}. The Chern-Simons formalism of AdS$_3$ is a powerful tool to explore the holographic aspects of the AdS$_3$ with various boundary conditions. In this paper, we apply the Chern-Simons formalism to study the charges of $T\bar{T}$ deformed CFT. We start from the Ba\~nados geometry, which is the most general AdS$_3$ solution with Brown-Henneaux boundary condition. The deformed Chern-Simons gauge connections were obtained through the field-dependent coordinate transformation. An analogy to the Ba\~nados geometry, we parametrize the deformed gauge connections by two independent functions, which corresponds to the $T\bar{T}$ deformed charges.
\par
After gauge fixing of the deformed gauge fields and imposing the mixed boundary condition, the residual gauge symmetries can be found. The left-moving gauge fields and the right-moving gauge fields are entangled. The residual gauge generators can be parametrized by $\epsilon=\lambda_{-1}-\mu\bar{\lambda}_{-1}$ and  $\bar{\epsilon}=\bar{\lambda}_{1}-\mu\lambda_{1}$. Then the variations of the charges give the algebra of the charges under the gauge transformation concerning $\epsilon,\bar{\epsilon}$. The resulting charge algebra is a non-linear deformation of the Virasoro algebra. We expand the Poisson bracket algebra of the charges perturbatively around $\mu=0$ and the leading order reproduces the Virasoro algebra. The first-order correction induces coupling between the deformed charges ${Q}$ and $\bar{Q}$.
\par
In~\cite{Guica:2019nzm}, the asymptotic symmetry of AdS$_3$ with the mixed boundary condition is described by two commuting copies of field-dependent Virasoro algebra. The different algebra structure was obtained in~\cite{Kraus:2021cwf} when they consider the asymptotic symmetry of the deformed metric on the finite surface $r=r_c$ with Dirichlet boundary condition. In~\cite{Guica:2020uhm}, it turns out that there are some uncertain winding terms in the deformed charge algebra. We show the difference between the two algebras offered by \cite{Guica:2019nzm} and \cite{Kraus:2021cwf} comes from the ambiguous definition of the deformed charges.

\section*{Acknowledgments}
We would like to thank Per Kraus, Pujian Mao, Yuan Sun, and Long Zhao for the helpful discussion. S.H. would like to appreciate the financial support from Jilin University, Max Planck Partner group as well as Natural Science Foundation of China Grants (No.12075101, No. 12047569). Y.G. would like to thank the support from the National Natural Science Foundation of China (NSFC) under Grant No.11875082, 12047503.

\appendix
\section{Solving the variation of the charges}
\label{app-A}
In this appendix, we treat in more detail how to solve the generator of the residual gauge transformation. We will also extract the variation of the charges under the gauge transformation. For convenience, we rewrite the equations as following
\begin{align}
\label{eq-lambda-1}
-\mu\delta_{\lambda}\bar q=&\lambda'_{-1}+\lambda_0(1-\mu \bar q),\\
\label{eq-lambda0}
0=&\lambda'_0+2(\lambda_{-1}q-\lambda_1(1-\mu \bar q)),\\
\label{eq-lambda1}
\delta_{\lambda} q=&\lambda'_1-\lambda_0q,\\
\label{eq-lambdabar-1}
-\delta_{\bar{\lambda}}\bar q=&\bar\lambda_{-1}'-\bar\lambda_0\bar q,\\
\label{eq-lambdabar0}
0=&\lambda'_0+2\Big(\bar\lambda_1\bar q-\bar\lambda_{-1}(1-\mu q)\Big),\\
\label{eq-lambdabar1}
\mu\delta_{\bar{\lambda}}q=&\bar\lambda'_1+\bar\lambda_0(1-\mu q),\\
\delta_{\lambda} q=&\delta_{\bar{\lambda}}q,\quad \delta_{\lambda}\bar{q}=\delta_{\bar{\lambda}}\bar{q}.
\end{align}
First of all, from~\eqref{eq-lambda-1},~\eqref{eq-lambda1},~\eqref{eq-lambdabar-1} and~\eqref{eq-lambdabar1}, by eliminating $\delta_{\lambda}q$ and $\delta_{\bar{\lambda}}\bar{q}$ we can get the equations for $\lambda_{0},\bar{\lambda}_{0}$
\begin{align}
\mu  \left(\bar{\lambda}_0(\theta )-\lambda _0(\theta )\right) \bar{q}(\theta )+\epsilon'(\theta )+\lambda _0(\theta )=0,\\
\bar{\epsilon}'(\theta )+\bar{\lambda}_0(\theta ) (1-\mu  q(\theta ))+\mu  \lambda _0(\theta ) q(\theta )=0,
\end{align}
Solving these equations, we obtain
\begin{align}
\label{lambda0}
\lambda _0(\theta )=&(\zeta (\theta )-1) \bar{\epsilon}'(\theta )-\zeta (\theta ) \epsilon'(\theta ),\\
\label{lambdabar0}
\bar{\lambda}_0(\theta )=& \left(\bar{\zeta }(\theta )-1\right) \epsilon'(\theta )-\bar{\zeta }(\theta ) \bar{\epsilon}'(\theta ),
\end{align}
where
\begin{align}
\zeta (\theta )=\frac{1-\mu  q}{1-\mu  \left(\bar{q}+q\right)},\quad \bar{\zeta }=\frac{1-\mu  \bar{q}}{1-\mu  \left(\bar{q}+q\right)}.
\end{align}
Substituting these solutions into~\eqref{eq-lambda0} and~\eqref{eq-lambdabar0}, the equations are following
\begin{align}
\frac{2 \lambda _{-1}(\theta ) \left(\bar{\zeta }(\theta )-1\right)-2 \mu  \lambda _1(\theta ) \bar{\zeta }(\theta )}{\mu  \left(\bar{\zeta }(\theta )+\zeta (\theta )-1\right)}+\zeta '(\theta ) \left(\bar{\epsilon}'(\theta )-\epsilon'(\theta )\right)+(\zeta (\theta )-1) \bar{\epsilon}''(\theta )-\zeta (\theta ) \epsilon''(\theta )=0,\\
\frac{2\bar{\lambda}_1(\theta )(\zeta (\theta )-1)-2 \mu\bar{\lambda}_{-1}(\theta )\zeta (\theta )}{\mu  \left(\bar{\zeta }(\theta )+\zeta (\theta )-1\right)}+\bar{\zeta }'(\theta )\left(\epsilon'(\theta )-\bar{\epsilon}'(\theta )\right) +\left(\bar{\zeta }(\theta )-1\right) \epsilon''(\theta )-\bar{\zeta }(\theta ) \bar{\epsilon}''(\theta )=0.
\end{align}
Combining with the definition of $\epsilon$ and $\bar{\epsilon}$,
\begin{align}
\epsilon(\theta )-(\lambda _{-1}(\theta )-\mu  \bar{\lambda}_{-1}(\theta ))=0,\quad \bar{\epsilon}(\theta )-(\bar{\lambda}_1(\theta )-\mu \lambda _1(\theta ))=0,
\end{align}
The solution implies one can express $\lambda_{-1},\lambda_{1},\bar\lambda_{-1},\bar\lambda_{1}$ in terms of the new parameters $\epsilon$ and $\bar{\epsilon}$, which read
\begin{align}
\lambda _{-1}(\theta )=&\frac{\zeta (\theta ) \bar{\zeta }(\theta )\epsilon(\theta )}{\bar{\zeta }(\theta )+\zeta (\theta )-1}+\frac{(\zeta (\theta )-1)\bar{\zeta }(\theta )\bar{\epsilon}(\theta )}{\bar{\zeta }(\theta )+\zeta (\theta )-1}\nonumber\\
&+\frac{1}{2} \mu\left(\bar{\zeta }(\theta ) \bar{\zeta }'(\theta )-(\zeta (\theta )-1) \zeta '(\theta )\right)\epsilon'(\theta )+\frac{1}{2} \mu   \left((\zeta (\theta )-1) \zeta '(\theta )-\bar{\zeta }(\theta ) \bar{\zeta }'(\theta )\right)\bar{\epsilon}'(\theta )\nonumber\\
&+\frac{1}{2} \mu  \left(\left(\bar{\zeta }(\theta )-1\right) \bar{\zeta }(\theta )-\zeta (\theta )^2+\zeta (\theta )\right)\epsilon''(\theta )+\frac{1}{2} \mu  \left((\zeta (\theta )-1)^2-\bar{\zeta }(\theta )^2\right) \bar{\epsilon}''(\theta ),
\end{align}
\begin{align}
\lambda_1(\theta )=&\frac{\epsilon(\theta ) \zeta (\theta ) \left(\bar{\zeta }(\theta )-1\right)}{\mu  \left(\bar{\zeta }(\theta )+\zeta (\theta )-1\right)}+\frac{(\zeta (\theta )-1) \bar{\epsilon}(\theta ) \left(\bar{\zeta }(\theta )-1\right)}{\mu  \left(\bar{\zeta }(\theta )+\zeta (\theta )-1\right)}\nonumber\\
&+\frac{1}{2} \epsilon'(\theta ) \left(\left(\bar{\zeta }(\theta )-1\right) \bar{\zeta }'(\theta )-\zeta (\theta ) \zeta '(\theta )\right)+\frac{1}{2} \bar{\epsilon}'(\theta ) \left(\zeta (\theta ) \zeta '(\theta )-\left(\bar{\zeta }(\theta )-1\right) \bar{\zeta }'(\theta )\right)\nonumber\\
&-\frac{1}{2} \left(\zeta (\theta )-\bar{\zeta }(\theta )\right) \left(\bar{\zeta }(\theta )+\zeta (\theta )-1\right) \bar{\epsilon}''(\theta )\nonumber\\
&-\frac{1}{2} \left(-\bar{\zeta }(\theta )+\zeta (\theta )+1\right) \left(\bar{\zeta }(\theta )+\zeta (\theta )-1\right)\epsilon''(\theta ),
\end{align}
\begin{align}
\bar{\lambda}_{-1}(\theta )=&\frac{\epsilon(\theta ) (\zeta (\theta )-1) \left(\bar{\zeta }(\theta )-1\right)}{\mu  \left(\bar{\zeta }(\theta )+\zeta (\theta )-1\right)}+\frac{(\zeta (\theta )-1) \bar{\epsilon}(\theta ) \bar{\zeta }(\theta )}{\mu  \left(\bar{\zeta }(\theta )+\zeta (\theta )-1\right)}\nonumber\\
&+\frac{1}{2}\epsilon'(\theta ) \left(\bar{\zeta }(\theta ) \bar{\zeta }'(\theta )-(\zeta (\theta )-1) \zeta '(\theta )\right)+\frac{1}{2} \bar{\epsilon}'(\theta ) \left((\zeta (\theta )-1) \zeta '(\theta )-\bar{\zeta }(\theta ) \bar{\zeta }'(\theta )\right)\nonumber\\
&+\frac{1}{2} \left(-\bar{\zeta }(\theta )+\zeta (\theta )-1\right) \left(\bar{\zeta }(\theta )+\zeta (\theta )-1\right) \bar{\epsilon}''(\theta )\nonumber\\
&-\frac{1}{2} \left(\zeta (\theta )-\bar{\zeta }(\theta )\right) \left(\bar{\zeta }(\theta )+\zeta (\theta )-1\right)\epsilon''(\theta ),
\end{align}
\begin{align}
\bar{\lambda}_1(\theta )=&\frac{\epsilon(\theta ) \zeta (\theta ) \left(\bar{\zeta }(\theta )-1\right)}{\bar{\zeta }(\theta )+\zeta (\theta )-1}+\frac{\zeta (\theta ) \bar{\epsilon}(\theta ) \bar{\zeta }(\theta )}{\bar{\zeta }(\theta )+\zeta (\theta )-1}\nonumber\\
&+\frac{1}{2} \epsilon'(\theta ) \left(\mu  \left(\bar{\zeta }(\theta )-1\right) \bar{\zeta }'(\theta )-\mu  \zeta (\theta ) \zeta '(\theta )\right)+\frac{1}{2} \mu  \bar{\epsilon}'(\theta ) \left(\zeta (\theta ) \zeta '(\theta )-\left(\bar{\zeta }(\theta )-1\right) \bar{\zeta }'(\theta )\right)\nonumber\\
&+\frac{1}{2} \mu  \left(\zeta (\theta )-\bar{\zeta }(\theta )\right) \left(\bar{\zeta }(\theta )+\zeta (\theta )-1\right) \bar{\epsilon}''(\theta )+\frac{1}{2} \mu  \left((\bar{\zeta}(\theta)-1)^2-\zeta(\theta )^2\right)\epsilon''(\theta ).
\end{align}
Finally, substituting back into~\eqref{eq-lambda1} and~\eqref{eq-lambdabar-1}, one can obtain
\begin{align}
\delta q=&\bar{\delta}q=\frac{\bar{\epsilon}(\theta ) \left((\bar{\zeta }(\theta )-1)\bar{\zeta }(\theta ) \zeta '(\theta )+(\zeta (\theta )-1) \zeta (\theta ) \bar{\zeta }'(\theta )\right)}{\mu  \left(\bar{\zeta }(\theta )+\zeta (\theta )-1\right)^2}\nonumber\\
&+\frac{\epsilon(\theta ) \left(\left(\bar{\zeta }(\theta )-1\right)^2 \zeta '(\theta )+\zeta (\theta )^2 \bar{\zeta }'(\theta )\right)}{\mu  \left(\bar{\zeta }(\theta )+\zeta (\theta )-1\right)^2}\nonumber\\
&+\frac{1}{2}\epsilon'(\theta ) \left(\frac{4 \zeta (\theta ) \left(\bar{\zeta }(\theta )-1\right)}{\mu  \left(\bar{\zeta }(\theta )+\zeta (\theta )-1\right)}+\bar{\zeta }'(\theta )^2+\left(\bar{\zeta }(\theta )-1\right) \bar{\zeta }''(\theta )-\zeta (\theta ) \zeta ''(\theta )-\zeta '(\theta )^2\right)\nonumber\\
&+\frac{1}{2} \bar{\epsilon}'(\theta ) \left(-\bar{\zeta }'(\theta )^2-\left(\bar{\zeta }(\theta )-1\right) \bar{\zeta }''(\theta )+\zeta (\theta ) \zeta ''(\theta )+\zeta '(\theta )^2\right)\nonumber\\
&-\frac{3}{2}\epsilon''(\theta ) \left(\zeta (\theta ) \zeta '(\theta )-\left(\bar{\zeta }(\theta )-1\right) \bar{\zeta }'(\theta )\right)+\frac{1}{2} \left(\left(2-3 \bar{\zeta }(\theta )\right) \bar{\zeta }'(\theta )+(3 \zeta (\theta )-1) \zeta '(\theta )\right) \bar{\epsilon}''(\theta )\nonumber\\
&-\frac{1}{2}\left(\bar{\zeta }(\theta )^2-\zeta (\theta )^2+(\zeta (\theta ))-\bar{\zeta }(\theta )\right) \bar{\epsilon}'''(\theta )-\frac{1}{2}\epsilon'''(\theta ) \left(\zeta (\theta )^2-(\bar{\zeta }(\theta )-2)\bar{\zeta }(\theta )-1\right),
\end{align}
\begin{align}
\delta \bar{q}=&\bar{\delta}\bar{q}=\frac{\epsilon(\theta ) \left(-(\bar{\zeta }(\theta )-1)\bar{\zeta }(\theta ) \zeta '(\theta )-(\zeta (\theta )-1) \zeta (\theta ) \bar{\zeta }'(\theta )\right)}{\mu  \left(\bar{\zeta }(\theta )+\zeta (\theta )-1\right)^2}\nonumber\\
&-\frac{\bar{\epsilon}(\theta ) \left(\bar{\zeta }(\theta )^2 \zeta '(\theta )+(\zeta (\theta )-1)^2 \bar{\zeta }'(\theta )\right)}{\mu  \left(\bar{\zeta }(\theta )+\zeta (\theta )-1\right)^2}\nonumber\\
&+\frac{1}{2}\epsilon'(\theta ) \left(-\bar{\zeta }'(\theta )^2-\bar{\zeta }(\theta ) \bar{\zeta }''(\theta )+\zeta (\theta ) \zeta ''(\theta )-\zeta ''(\theta )+\zeta '(\theta )^2\right)\nonumber\\
&+\frac{1}{2} \bar{\epsilon}'(\theta ) \left(-\frac{4 (\zeta (\theta )-1) \bar{\zeta }(\theta )}{\mu  \left(\bar{\zeta }(\theta )+\zeta (\theta )-1\right)}+\bar{\zeta }'(\theta )^2+\bar{\zeta }(\theta ) \bar{\zeta }''(\theta )-\zeta (\theta ) \zeta ''(\theta )+\zeta ''(\theta )-\zeta '(\theta )^2\right)\nonumber\\
&+\frac{1}{2} \epsilon''(\theta ) \left(\left(1-3 \bar{\zeta }(\theta )\right) \bar{\zeta }'(\theta )+(3 \zeta (\theta )-2) \zeta '(\theta )\right)+\frac{3}{2} \left(\bar{\zeta }(\theta ) \bar{\zeta }'(\theta )- (\zeta (\theta )-1) \zeta '(\theta )\right) \bar{\epsilon}''(\theta )\nonumber\\
&+\frac{1}{2}\epsilon'''(\theta ) \left(\zeta (\theta )^2-\bar{\zeta }(\theta )^2+\bar{\zeta }(\theta )-\zeta (\theta )\right)+\frac{1}{2} \left(\bar{\zeta }(\theta )^2-(\zeta (\theta )-2) \zeta (\theta )-1\right) \bar{\epsilon}'''(\theta )
\end{align}
Fortunately, it is convenient to introduce the auxiliary variables
\begin{align}
A=&\bar{A}=\frac{(\zeta (\theta )-1) \left(\bar{\zeta }(\theta )-1\right)}{\mu  \left(\bar{\zeta }(\theta )+\zeta (\theta )-1\right)}=\frac{\mu q \bar{q}}{1-\mu\left(\bar{q}+q\right)},\\
B=&\frac{1}{2}\Big(\zeta(\theta )^2-(\bar{\zeta}(\theta )-2) \bar{\zeta} (\theta )-2\Big)=\frac{\mu\left(2\bar{q}-\mu(q+\bar q)^2\right)}{2(1-\mu(q+\bar q))^2},\\
\bar B=&\frac{1}{2}\Big(\bar{\zeta }(\theta )^2-(\zeta (\theta )-2) \zeta (\theta )-2\Big)=\frac{\mu\left(2q-\mu(q+\bar q)^2\right)}{2(1-\mu(q+\bar q))^2},\\
C=&\frac{\zeta (\theta ) \left(\bar{\zeta }(\theta )-1\right)}{\mu  \left(\bar{\zeta }(\theta )+\zeta (\theta )-1\right)}=\frac{(1-\mu q)q}{1-\mu(q+\bar q)},\\
\bar C=&\frac{(\zeta (\theta )-1) \bar{\zeta }(\theta )}{\mu  \left(\bar{\zeta }(\theta )+\zeta (\theta )-1\right)}=\frac{(1-\mu \bar q)\bar q}{1-\mu(q+\bar q)},\\
D=&-\bar{D}=\bar{\zeta }(\theta )^2-\bar{\zeta }(\theta )-\zeta (\theta )^2+\zeta (\theta )=\frac{\mu  \left(q-\bar{q}\right)}{\left(1-\mu\left(\bar{q}+q\right)\right)^2},
\end{align}
Then
\begin{align}
A'=&\bar{A}'=\frac{\left(\bar{\zeta }(\theta )-1\right) \bar{\zeta }(\theta ) \zeta '(\theta )+(\zeta (\theta )-1) \zeta (\theta ) \bar{\zeta }'(\theta )}{\mu  \left(\bar{\zeta }(\theta )+\zeta (\theta )-1\right)^2},
\end{align}
\begin{align}
B'=&\zeta (\theta ) \zeta '(\theta )-\left(\bar{\zeta }(\theta )-1\right) \bar{\zeta }'(\theta ),\\
B''=&\zeta '(\theta )^2+\zeta (\theta ) \zeta ''(\theta )-\left(\bar{\zeta }(\theta )-1\right) \bar{\zeta }''(\theta )-\bar{\zeta }'(\theta )^2,
\end{align}
\begin{align}
\bar{B}'=&\bar{\zeta }(\theta ) \bar{\zeta }'(\theta )-(\zeta (\theta )-1) \zeta '(\theta ),\\
\bar{B}''=&\bar{\zeta }'(\theta )^2+\bar{\zeta }(\theta ) \bar{\zeta }''(\theta )-(\zeta (\theta )-1) \zeta ''(\theta )-\zeta '(\theta )^2,
\end{align}
\begin{align}
C'=&\frac{\left(\bar{\zeta }(\theta )-1\right)^2 \zeta '(\theta )+\zeta (\theta )^2 \bar{\zeta }'(\theta )}{\mu  \left(\bar{\zeta }(\theta )+\zeta (\theta )-1\right)^2},
\end{align}
\begin{align}
\bar{C}'=&\frac{\bar{\zeta }(\theta )^2 \zeta '(\theta )+(\zeta (\theta )-1)^2 \bar{\zeta }'(\theta )}{\mu  \left(\bar{\zeta }(\theta )+\zeta (\theta )-1\right)^2},
\end{align}
\begin{align}
D'=&-\bar{D}'=\left(2 \bar{\zeta }(\theta )-1\right) \bar{\zeta }'(\theta )+(1-2\zeta (\theta )) \zeta '(\theta ).
\end{align}
Finally, $\delta_{\lambda}q$ and $\delta_{\bar{\lambda}}\bar{q}$ can be formulated as
\begin{align}
\delta_{\lambda} q=\delta_{\bar{\lambda}}q=&\epsilon(\theta)C'+\bar{\epsilon}(\theta)A'+\frac{1}{2}\bar{\epsilon}'(\theta)B''-\frac{1}{2}\epsilon'(\theta)(B''-4C)\nonumber\\
&-\frac{1}{2}\bar{\epsilon}''(\theta)(D'-B')-\frac{3}{2}\epsilon''(\theta)B'-\frac{1}{2}\bar{\epsilon}'''(\theta )D-\frac{1}{2}\epsilon'''(\theta )(2B+1),\\
\delta_{\lambda}\bar{q}=\delta_{\bar{\lambda}}\bar{q}=&-\epsilon(\theta)\bar{A}'-\bar{\epsilon}(\theta)\bar{C}'-\frac{1}{2}\epsilon'(\theta)\bar{B}''+\frac{1}{2}\bar{\epsilon}'(\theta)(\bar{B}''-4\bar{C})\nonumber\\
&+\frac{1}{2}\epsilon''(\theta)(\bar{D}'-\bar{B}')+\frac{3}{2}\bar{\epsilon}''(\theta)\bar{B}'+\frac{1}{2}\epsilon'''(\theta )\bar{D}+\frac{1}{2}\bar{\epsilon}'''(\theta )(2\bar{B}+1).
\end{align}
\section{Constraints on the gauge transformation}
\label{app-B}
In this appendix, we would like to derive the evolution equation of the parameters $\epsilon,\epsilon'$ from the following equations
\begin{align}
\label{cons_lambda-1}
\delta_{\lambda}\left(K(1-\mu\bar{q})\right)=&\partial_{t}\lambda_{-1}+K\lambda_{0}(1-\mu\bar{q}),\\
\label{cons_lambda0}
0=&\partial_{t}\lambda_{0}+2K\left(\lambda_{-1}q-\lambda_{1}(1-\mu\bar{q})\right),\\
\label{cons_lambda1}
\delta_{\lambda}\left(Kq\right)=&\partial_{t}\lambda_{1}-K\lambda_{0}q,\\
\label{cons_lambdabar-1}
-\delta_{\bar{\lambda}}(\bar{K}\bar{q})=&\partial_{t}\bar{\lambda}_{-1}-\bar{K}\bar{\lambda}_{0}\bar{q},\\
\label{cons_lambdabar0}
0=&\partial_{t}\bar{\lambda}_{0}+2\bar{K}\left(\bar{\lambda}_{1}\bar{q}-\bar{\lambda}_{-1}(1-\mu q)\right),\\
\label{cons_lambdabar1}
-\delta_{\bar{\lambda}}\left(\bar{K}(1-\mu q)\right)=&\partial_{t}\bar{\lambda}_{1}+\bar{K}\bar{\lambda}_{0}(1-\mu q).
\end{align}
Firstly, by using the definition of $K,\bar{K}$, one can find the relations
\begin{align}
K(1-\mu\bar{q})+\mu\bar{K}\bar{q}=1,\quad \mu Kq+\bar{K}(1-\mu{q})=1
\end{align}
then
\begin{align}
\label{KKbar-variation-relation1}
\delta_{\lambda}\left(K(1-\mu\bar{q})\right)+\mu\delta_{\bar{\lambda}}(\bar{K}\bar{q})=0,\\
\label{KKbar-variation-relation2}
\delta_{\bar\lambda}\left(\bar{K}(1-\mu{q})\right)+\mu\delta_{\lambda}(Kq)=0.
\end{align}
Combining~\eqref{cons_lambda-1},~\eqref{cons_lambdabar-1},~\eqref{KKbar-variation-relation1}, ~\eqref{variation_relation_1}, and~\eqref{variation_relation_2}, one can obtain
\begin{align}
\label{pt-epsilon}
\partial_{t}\epsilon=K(1-\mu\bar{q})\lambda_0+\mu\bar{K}\bar{q}\bar{\lambda}_0.
\end{align}
From~\eqref{cons_lambda1} and~\eqref{cons_lambdabar1}, one can get
\begin{align}
\label{pt-epsilonbar}
\partial_{t}\bar{\epsilon}=\bar{K}(1-\mu q)\bar{\lambda}_0+\mu Kq\lambda_0.
\end{align}
Finally, plugging~\eqref{lambda0} and~\eqref{lambdabar0} into~\eqref{pt-epsilon} and~\eqref{pt-epsilonbar}, one can arrive at
\begin{align}
\label{B12}
-\partial_{t}\epsilon=&\epsilon'-\frac{2 \mu  \bar{q} \left(1-\mu  \bar{q}\right)}{\left(1-\mu \left(\bar{q}+q\right)\right)^2}
\left(\epsilon'+\bar{\epsilon}'\right),\\
\label{B13}
\partial_{t}\bar{\epsilon}=&\bar{\epsilon}'-\frac{2 \mu q\left(1-\mu  q\right)}{\left(1-\mu  \left(\bar{q}+q\right)\right)^2}\left(\epsilon'+\bar{\epsilon}'\right).
\end{align}
In addition, from~\eqref{eq-lambda0},~\eqref{eq-lambdabar0},\eqref{cons_lambda0} and ~\eqref{cons_lambdabar0}, the $\lambda_{0}$ and $\bar{\lambda}_{0}$ obey
\begin{align}
\partial_{t}\lambda_{0}=&\frac{1+\mu(\bar q-q)}{1-\mu(q+\bar q)}\lambda_{0}',\\
\partial_{t}\lambda_{0}=&-\frac{1-\mu(\bar q-q)}{1-\mu(q+\bar q)}\bar{\lambda}_{0}'.
\end{align}
%%%%%%%%%%%%%%%%%%%%%%%%%%%%%%%%
%  \bibliographystyle{JHEP}
%  \bibliography{reference}

\begin{thebibliography}{10}

\bibitem{Smirnov:2016lqw}
F.~A. Smirnov and A.~B. Zamolodchikov, \emph{{On space of integrable quantum
  field theories}},
  \href{https://doi.org/10.1016/j.nuclphysb.2016.12.014}{\emph{Nucl. Phys. B}
  {\bfseries 915} (2017) 363}
  [\href{https://arxiv.org/abs/1608.05499}{{\ttfamily 1608.05499}}].

\bibitem{Cavaglia:2016oda}
A.~Cavagli\`a, S.~Negro, I.~M. Sz\'ecs\'enyi and R.~Tateo, \emph{{$T
  \bar{T}$-deformed 2D Quantum Field Theories}},
  \href{https://doi.org/10.1007/JHEP10(2016)112}{\emph{JHEP} {\bfseries 10}
  (2016) 112} [\href{https://arxiv.org/abs/1608.05534}{{\ttfamily
  1608.05534}}].

\bibitem{McGough:2016lol}
L.~McGough, M.~Mezei and H.~Verlinde, \emph{{Moving the CFT into the bulk with
  $ T\overline{T} $}},
  \href{https://doi.org/10.1007/JHEP04(2018)010}{\emph{JHEP} {\bfseries 04}
  (2018) 010} [\href{https://arxiv.org/abs/1611.03470}{{\ttfamily
  1611.03470}}].

\bibitem{Bonelli:2018kik}
G.~Bonelli, N.~Doroud and M.~Zhu, \emph{{$T \bar{T}$-deformations in closed
  form}}, \href{https://doi.org/10.1007/JHEP06(2018)149}{\emph{JHEP} {\bfseries
  06} (2018) 149} [\href{https://arxiv.org/abs/1804.10967}{{\ttfamily
  1804.10967}}].

\bibitem{Datta:2018thy}
S.~Datta and Y.~Jiang, \emph{{$T\bar{T}$ deformed partition functions}},
  \href{https://doi.org/10.1007/JHEP08(2018)106}{\emph{JHEP} {\bfseries 08}
  (2018) 106} [\href{https://arxiv.org/abs/1806.07426}{{\ttfamily
  1806.07426}}].

\bibitem{Aharony:2018bad}
O.~Aharony, S.~Datta, A.~Giveon, Y.~Jiang and D.~Kutasov, \emph{{Modular
  invariance and uniqueness of $T\bar{T}$ deformed CFT}},
  \href{https://doi.org/10.1007/JHEP01(2019)086}{\emph{JHEP} {\bfseries 01}
  (2019) 086} [\href{https://arxiv.org/abs/1808.02492}{{\ttfamily
  1808.02492}}].

\bibitem{Dubovsky:2017cnj}
S.~Dubovsky, V.~Gorbenko and M.~Mirbabayi, \emph{{Asymptotic fragility, near
  AdS$_{2}$ holography and $ T\overline{T} $}},
  \href{https://doi.org/10.1007/JHEP09(2017)136}{\emph{JHEP} {\bfseries 09}
  (2017) 136} [\href{https://arxiv.org/abs/1706.06604}{{\ttfamily
  1706.06604}}].

\bibitem{Dubovsky:2018bmo}
S.~Dubovsky, V.~Gorbenko and G.~Hern\'andez-Chifflet, \emph{{$ T\overline{T} $
  partition function from topological gravity}},
  \href{https://doi.org/10.1007/JHEP09(2018)158}{\emph{JHEP} {\bfseries 09}
  (2018) 158} [\href{https://arxiv.org/abs/1805.07386}{{\ttfamily
  1805.07386}}].

\bibitem{Ishii:2019uwk}
T.~Ishii, S.~Okumura, J.-I. Sakamoto and K.~Yoshida, \emph{{Gravitational
  perturbations as $T\bar{T}$-deformations in 2D dilaton gravity systems}},
  \href{https://doi.org/10.1016/j.nuclphysb.2019.114901}{\emph{Nucl. Phys. B}
  {\bfseries 951} (2020) 114901}
  [\href{https://arxiv.org/abs/1906.03865}{{\ttfamily 1906.03865}}].

\bibitem{Cardy:2018sdv}
J.~Cardy, \emph{{The $ T\overline{T} $ deformation of quantum field theory as
  random geometry}}, \href{https://doi.org/10.1007/JHEP10(2018)186}{\emph{JHEP}
  {\bfseries 10} (2018) 186}
  [\href{https://arxiv.org/abs/1801.06895}{{\ttfamily 1801.06895}}].


\bibitem{Conti:2018tca}
R.~Conti, S.~Negro and R.~Tateo, \emph{{The $ \mathrm{T}\overline{\mathrm{T}} $
  perturbation and its geometric interpretation}},
  \href{https://doi.org/10.1007/JHEP02(2019)085}{\emph{JHEP} {\bfseries 02}
  (2019) 085} [\href{https://arxiv.org/abs/1809.09593}{{\ttfamily
  1809.09593}}].

\bibitem{Conti:2019dxg}
R.~Conti, S.~Negro and R.~Tateo, \emph{{Conserved currents and
  $\text{T}\bar{\text{T}}_s$ irrelevant deformations of 2D integrable field
  theories}}, \href{https://doi.org/10.1007/JHEP11(2019)120}{\emph{JHEP}
  {\bfseries 11} (2019) 120}
  [\href{https://arxiv.org/abs/1904.09141}{{\ttfamily 1904.09141}}].

\bibitem{Callebaut:2019omt}
N.~Callebaut, J.~Kruthoff and H.~Verlinde, \emph{{$ T\overline{T} $ deformed
  CFT as a non-critical string}},
  \href{https://doi.org/10.1007/JHEP04(2020)084}{\emph{JHEP} {\bfseries 04}
  (2020) 084} [\href{https://arxiv.org/abs/1910.13578}{{\ttfamily
  1910.13578}}].

\bibitem{Tolley:2019nmm}
A.~J. Tolley, \emph{{$ T\overline{T} $ deformations, massive gravity and
  non-critical strings}},
  \href{https://doi.org/10.1007/JHEP06(2020)050}{\emph{JHEP} {\bfseries 06}
  (2020) 050} [\href{https://arxiv.org/abs/1911.06142}{{\ttfamily
  1911.06142}}].

\bibitem{Kraus:2018xrn}
P.~Kraus, J.~Liu and D.~Marolf, \emph{{Cutoff AdS$_{3}$ versus the $
  T\overline{T} $ deformation}},
  \href{https://doi.org/10.1007/JHEP07(2018)027}{\emph{JHEP} {\bfseries 07}
  (2018) 027} [\href{https://arxiv.org/abs/1801.02714}{{\ttfamily
  1801.02714}}].

\bibitem{Shyam:2017znq}
V.~Shyam, \emph{{Background independent holographic dual to $T\bar{T}$ deformed
  CFT with large central charge in 2 dimensions}},
  \href{https://doi.org/10.1007/JHEP10(2017)108}{\emph{JHEP} {\bfseries 10}
  (2017) 108} [\href{https://arxiv.org/abs/1707.08118}{{\ttfamily
  1707.08118}}].

\bibitem{Donnelly:2019pie}
W.~Donnelly, E.~LePage, Y.-Y. Li, A.~Pereira and V.~Shyam, \emph{{Quantum
  corrections to finite radius holography and holographic entanglement
  entropy}}, \href{https://doi.org/10.1007/JHEP05(2020)006}{\emph{JHEP}
  {\bfseries 05} (2020) 006}
  [\href{https://arxiv.org/abs/1909.11402}{{\ttfamily 1909.11402}}].

\bibitem{Klebanov:1999tb}
I.~R. Klebanov and E.~Witten, \emph{{AdS / CFT correspondence and symmetry
  breaking}}, \href{https://doi.org/10.1016/S0550-3213(99)00387-9}{\emph{Nucl.
  Phys. B} {\bfseries 556} (1999) 89}
  [\href{https://arxiv.org/abs/hep-th/9905104}{{\ttfamily hep-th/9905104}}].

\bibitem{Witten:2001ua}
E.~Witten, \emph{{Multitrace operators, boundary conditions, and AdS / CFT
  correspondence}},  \href{https://arxiv.org/abs/hep-th/0112258}{{\ttfamily
  hep-th/0112258}}.

\bibitem{Papadimitriou:2007sj}
I.~Papadimitriou, \emph{{Multi-Trace Deformations in AdS/CFT: Exploring the
  Vacuum Structure of the Deformed CFT}},
  \href{https://doi.org/10.1088/1126-6708/2007/05/075}{\emph{JHEP} {\bfseries
  05} (2007) 075} [\href{https://arxiv.org/abs/hep-th/0703152}{{\ttfamily
  hep-th/0703152}}].

\bibitem{Guica:2019nzm}
M.~Guica and R.~Monten, \emph{{$T\bar T$ and the mirage of a bulk cutoff}},
  \href{https://doi.org/10.21468/SciPostPhys.10.2.024}{\emph{SciPost Phys.}
  {\bfseries 10} (2021) 024}
  [\href{https://arxiv.org/abs/1906.11251}{{\ttfamily 1906.11251}}].

\bibitem{Ouyang:2020rpq}
H.~Ouyang and H.~Shu, \emph{{$T\bar{T}$ deformation of chiral bosons and
  Chern\textendash{}Simons $\hbox {AdS}_3$ gravity}},
  \href{https://doi.org/10.1140/epjc/s10052-020-08738-6}{\emph{Eur. Phys. J. C}
  {\bfseries 80} (2020) 1155}
  [\href{https://arxiv.org/abs/2006.10514}{{\ttfamily 2006.10514}}].

\bibitem{He:2020hhm}
M.~He and Y.-h. Gao, \emph{{$T\bar{T}/J\bar{T}$-deformed WZW models from
  Chern-Simons AdS$_3$ gravity with mixed boundary conditions}},
  \href{https://doi.org/10.1103/PhysRevD.103.126019}{\emph{Phys. Rev. D}
  {\bfseries 103} (2021) 126019}
  [\href{https://arxiv.org/abs/2012.05726}{{\ttfamily 2012.05726}}].

\bibitem{Giribet:2017imm}
G.~Giribet, \emph{{$T\bar{T}$-deformations, AdS/CFT and correlation
  functions}}, \href{https://doi.org/10.1007/JHEP02(2018)114}{\emph{JHEP}
  {\bfseries 02} (2018) 114}
  [\href{https://arxiv.org/abs/1711.02716}{{\ttfamily 1711.02716}}].

\bibitem{Donnelly:2018bef}
W.~Donnelly and V.~Shyam, \emph{{Entanglement entropy and $T \overline{T}$
  deformation}},
  \href{https://doi.org/10.1103/PhysRevLett.121.131602}{\emph{Phys. Rev. Lett.}
  {\bfseries 121} (2018) 131602}
  [\href{https://arxiv.org/abs/1806.07444}{{\ttfamily 1806.07444}}].

\bibitem{Chen:2018eqk}
B.~Chen, L.~Chen and P.-X. Hao, \emph{{Entanglement entropy in
  $T\overline{T}$-deformed CFT}},
  \href{https://doi.org/10.1103/PhysRevD.98.086025}{\emph{Phys. Rev. D}
  {\bfseries 98} (2018) 086025}
  [\href{https://arxiv.org/abs/1807.08293}{{\ttfamily 1807.08293}}].

\bibitem{Jeong:2019ylz}
H.-S. Jeong, K.-Y. Kim and M.~Nishida, \emph{{Entanglement and R\'enyi entropy
  of multiple intervals in $T\overline{T}$-deformed CFT and holography}},
  \href{https://doi.org/10.1103/PhysRevD.100.106015}{\emph{Phys. Rev. D}
  {\bfseries 100} (2019) 106015}
  [\href{https://arxiv.org/abs/1906.03894}{{\ttfamily 1906.03894}}].

\bibitem{Grieninger:2019zts}
S.~Grieninger, \emph{{Entanglement entropy and $ T\overline{T} $ deformations
  beyond antipodal points from holography}},
  \href{https://doi.org/10.1007/JHEP11(2019)171}{\emph{JHEP} {\bfseries 11}
  (2019) 171} [\href{https://arxiv.org/abs/1908.10372}{{\ttfamily
  1908.10372}}].

\bibitem{Jafari:2019qns}
G.~Jafari, A.~Naseh and H.~Zolfi, \emph{{Path Integral Optimization for
  $T\bar{T}$ Deformation}},
  \href{https://doi.org/10.1103/PhysRevD.101.026007}{\emph{Phys. Rev. D}
  {\bfseries 101} (2020) 026007}
  [\href{https://arxiv.org/abs/1909.02357}{{\ttfamily 1909.02357}}].

\bibitem{Chen:2019mis}
B.~Chen, L.~Chen and C.-Y. Zhang, \emph{{Surface/state correspondence and
  $T\overline{T}$ deformation}},
  \href{https://doi.org/10.1103/PhysRevD.101.106011}{\emph{Phys. Rev. D}
  {\bfseries 101} (2020) 106011}
  [\href{https://arxiv.org/abs/1907.12110}{{\ttfamily 1907.12110}}].

\bibitem{Mazenc:2019cfg}
E.~A. Mazenc, V.~Shyam and R.~M. Soni, \emph{{A $T \bar{T}$ Deformation for
  Curved Spacetimes from 3d Gravity}},
  \href{https://arxiv.org/abs/1912.09179}{{\ttfamily 1912.09179}}.

\bibitem{Li:2020pwa}
Y.~Li and Y.~Zhou, \emph{{Cutoff AdS$_{3}$ versus $ T\overline{T} $ CFT$_{2}$
  in the large central charge sector: correlators of energy-momentum tensor}},
  \href{https://doi.org/10.1007/JHEP12(2020)168}{\emph{JHEP} {\bfseries 12}
  (2020) 168} [\href{https://arxiv.org/abs/2005.01693}{{\ttfamily
  2005.01693}}].

\bibitem{Li:2020zjb}
Y.~Li, \emph{{Comments on large central charge $T\bar{T}$ deformed conformal
  field theory and cutoff AdS holography}},
  \href{https://arxiv.org/abs/2012.14414}{{\ttfamily 2012.14414}}.

\bibitem{Caputa:2020lpa}
P.~Caputa, S.~Datta, Y.~Jiang and P.~Kraus, \emph{{Geometrizing $ T\overline{T}
  $}}, \href{https://doi.org/10.1007/JHEP03(2021)140}{\emph{JHEP} {\bfseries
  03} (2021) 140} [\href{https://arxiv.org/abs/2011.04664}{{\ttfamily
  2011.04664}}].

\bibitem{Hirano:2020ppu}
S.~Hirano, T.~Nakajima and M.~Shigemori, \emph{{$ T\overline{T} $ Deformation
  of stress-tensor correlators from random geometry}},
  \href{https://doi.org/10.1007/JHEP04(2021)270}{\emph{JHEP} {\bfseries 04}
  (2021) 270} [\href{https://arxiv.org/abs/2012.03972}{{\ttfamily
  2012.03972}}].

\bibitem{Araujo:2018rho}
T.~Araujo, E.~O. Colg\'ain, Y.~Sakatani, M.~M. Sheikh-Jabbari and
  H.~Yavartanoo, \emph{{Holographic integration of $T \bar{T}$
  \textbackslash{}\& $J \bar{T}$ via $O(d,d)$}},
  \href{https://doi.org/10.1007/JHEP03(2019)168}{\emph{JHEP} {\bfseries 03}
  (2019) 168} [\href{https://arxiv.org/abs/1811.03050}{{\ttfamily
  1811.03050}}].

\bibitem{Babaei-Aghbolagh:2020kjg}
H.~Babaei-Aghbolagh, K.~B. Velni, D.~M. Yekta and H.~Mohammadzadeh, \emph{{$
  T\overline{T} $-like flows in non-linear electrodynamic theories and
  S-duality}}, \href{https://doi.org/10.1007/JHEP04(2021)187}{\emph{JHEP}
  {\bfseries 04} (2021) 187}
  [\href{https://arxiv.org/abs/2012.13636}{{\ttfamily 2012.13636}}].

\bibitem{Jiang:2019epa}
Y.~Jiang, \emph{{A pedagogical review on solvable irrelevant deformations of 2D
  quantum field theory}},
  \href{https://doi.org/10.1088/1572-9494/abe4c9}{\emph{Commun. Theor. Phys.}
  {\bfseries 73} (2021) 057201}
  [\href{https://arxiv.org/abs/1904.13376}{{\ttfamily 1904.13376}}].

\bibitem{LeFloch:2019wlf}
B.~Le~Floch and M.~Mezei, \emph{{KdV charges in $T\bar{T}$ theories and new
  models with super-Hagedorn behavior}},
  \href{https://doi.org/10.21468/SciPostPhys.7.4.043}{\emph{SciPost Phys.}
  {\bfseries 7} (2019) 043} [\href{https://arxiv.org/abs/1907.02516}{{\ttfamily
  1907.02516}}].

\bibitem{Cardy:2019qao}
J.~Cardy, \emph{{$T\bar T$ deformation of correlation functions}},
  \href{https://doi.org/10.1007/JHEP12(2019)160}{\emph{JHEP} {\bfseries 12}
  (2019) 160} [\href{https://arxiv.org/abs/1907.03394}{{\ttfamily
  1907.03394}}].

\bibitem{He:2019vzf}
S.~He and H.~Shu, \emph{{Correlation functions, entanglement and chaos in the $
  T\overline{T}/J\overline{T} $-deformed CFTs}},
  \href{https://doi.org/10.1007/JHEP02(2020)088}{\emph{JHEP} {\bfseries 02}
  (2020) 088} [\href{https://arxiv.org/abs/1907.12603}{{\ttfamily
  1907.12603}}].

\bibitem{He:2019ahx}
S.~He, J.-R. Sun and Y.~Sun, \emph{{The correlation function of (1,1) and (2,2)
  supersymmetric theories with $T\bar{T}$ deformation}},
  \href{https://doi.org/10.1007/JHEP04(2020)100}{\emph{JHEP} {\bfseries 04}
  (2020) 100} [\href{https://arxiv.org/abs/1912.11461}{{\ttfamily
  1912.11461}}].

\bibitem{Kruthoff:2020hsi}
J.~Kruthoff and O.~Parrikar, \emph{{On the flow of states under
  $T\overline{T}$}},  \href{https://arxiv.org/abs/2006.03054}{{\ttfamily
  2006.03054}}.

\bibitem{He:2020udl}
S.~He and Y.~Sun, \emph{{Correlation functions of CFTs on a torus with a
  $T\overline{T}$ deformation}},
  \href{https://doi.org/10.1103/PhysRevD.102.026023}{\emph{Phys. Rev. D}
  {\bfseries 102} (2020) 026023}
  [\href{https://arxiv.org/abs/2004.07486}{{\ttfamily 2004.07486}}].

\bibitem{He:2020cxp}
S.~He, Y.~Sun and Y.-X. Zhang, \emph{{$T\bar{T}$-flow effects on torus
  partition functions}},  \href{https://arxiv.org/abs/2011.02902}{{\ttfamily
  2011.02902}}.

\bibitem{He:2020qcs}
S.~He, \emph{{Note on higher-point correlation functions of the $T\bar T$ or
  $J\bar T$ deformed CFTs}},
  \href{https://doi.org/10.1007/s11433-021-1741-1}{\emph{Sci. China Phys. Mech.
  Astron.} {\bfseries 64} (2021) 291011}
  [\href{https://arxiv.org/abs/2012.06202}{{\ttfamily 2012.06202}}].

\bibitem{Cardy:2018jho}
J.~Cardy, \emph{{$T\overline T$ deformations of non-Lorentz invariant field
  theories}},  \href{https://arxiv.org/abs/1809.07849}{{\ttfamily 1809.07849}}.

\bibitem{Marchetto:2019yyt}
E.~Marchetto, A.~Sfondrini and Z.~Yang, \emph{{$T\bar{T}$ Deformations and
  Integrable Spin Chains}},
  \href{https://doi.org/10.1103/PhysRevLett.124.100601}{\emph{Phys. Rev. Lett.}
  {\bfseries 124} (2020) 100601}
  [\href{https://arxiv.org/abs/1911.12315}{{\ttfamily 1911.12315}}].

\bibitem{Cardy:2020olv}
J.~Cardy and B.~Doyon, \emph{{$T{\overline T}$ deformations and the width of
  fundamental particles}},  \href{https://arxiv.org/abs/2010.15733}{{\ttfamily
  2010.15733}}.

\bibitem{Medenjak:2020ppv}
M.~Medenjak, G.~Policastro and T.~Yoshimura, \emph{{$T\bar{T}$-Deformed
  Conformal Field Theories out of Equilibrium}},
  \href{https://doi.org/10.1103/PhysRevLett.126.121601}{\emph{Phys. Rev. Lett.}
  {\bfseries 126} (2021) 121601}
  [\href{https://arxiv.org/abs/2011.05827}{{\ttfamily 2011.05827}}].

\bibitem{Jiang:2020nnb}
Y.~Jiang, \emph{{$\mathrm{T}\overline{\mathrm{T}}$-deformed 1d Bose gas}},
  \href{https://arxiv.org/abs/2011.00637}{{\ttfamily 2011.00637}}.

\bibitem{Chen:2020jdi}
B.~Chen, J.~Hou and J.~Tian, \emph{{Note on the nonrelativistic
  TT\textasciimacron{} deformation}},
  \href{https://doi.org/10.1103/PhysRevD.104.025004}{\emph{Phys. Rev. D}
  {\bfseries 104} (2021) 025004}
  [\href{https://arxiv.org/abs/2012.14091}{{\ttfamily 2012.14091}}].

\bibitem{Jorjadze:2020ili}
G.~Jorjadze and S.~Theisen, \emph{{Canonical maps and integrability in $T\bar
  T$ deformed 2d CFTs}},  \href{https://arxiv.org/abs/2001.03563}{{\ttfamily
  2001.03563}}.

\bibitem{Guica:2020uhm}
M.~Guica and R.~Monten, \emph{{Infinite pseudo-conformal symmetries of
  classical $T \bar T$, $J \bar T $ and $J T_a$ - deformed CFTs}},
  \href{https://arxiv.org/abs/2011.05445}{{\ttfamily 2011.05445}}.

\bibitem{Kraus:2021cwf}
P.~Kraus, R.~Monten and R.~M. Myers, \emph{{3D Gravity in a Box}},
  \href{https://arxiv.org/abs/2103.13398}{{\ttfamily 2103.13398}}.

\bibitem{Brown:1986nw}
J.~D. Brown and M.~Henneaux, \emph{{Central Charges in the Canonical
  Realization of Asymptotic Symmetries: An Example from Three-Dimensional
  Gravity}}, \href{https://doi.org/10.1007/BF01211590}{\emph{Commun. Math.
  Phys.} {\bfseries 104} (1986) 207}.

\bibitem{Banados:1998gg}
M.~Banados, \emph{{Three-dimensional quantum geometry and black holes}},
  \href{https://doi.org/10.1063/1.59661}{\emph{AIP Conf. Proc.} {\bfseries 484}
  (1999) 147} [\href{https://arxiv.org/abs/hep-th/9901148}{{\ttfamily
  hep-th/9901148}}].

\bibitem{Banados:1998ta}
M.~Banados, T.~Brotz and M.~E. Ortiz, \emph{{Boundary dynamics and the
  statistical mechanics of the (2+1)-dimensional black hole}},
  \href{https://doi.org/10.1016/S0550-3213(99)00069-3}{\emph{Nucl. Phys. B}
  {\bfseries 545} (1999) 340}
  [\href{https://arxiv.org/abs/hep-th/9802076}{{\ttfamily hep-th/9802076}}].

\bibitem{Cotler:2018zff}
J.~Cotler and K.~Jensen, \emph{{A theory of reparameterizations for AdS$_3$
  gravity}}, \href{https://doi.org/10.1007/JHEP02(2019)079}{\emph{JHEP}
  {\bfseries 02} (2019) 079}
  [\href{https://arxiv.org/abs/1808.03263}{{\ttfamily 1808.03263}}].

\bibitem{Henneaux:2010xg}
M.~Henneaux and S.-J. Rey, \emph{{Nonlinear $W_{infinity}$ as Asymptotic
  Symmetry of Three-Dimensional Higher Spin Anti-de Sitter Gravity}},
  \href{https://doi.org/10.1007/JHEP12(2010)007}{\emph{JHEP} {\bfseries 12}
  (2010) 007} [\href{https://arxiv.org/abs/1008.4579}{{\ttfamily 1008.4579}}].

\bibitem{Campoleoni:2010zq}
A.~Campoleoni, S.~Fredenhagen, S.~Pfenninger and S.~Theisen, \emph{{Asymptotic
  symmetries of three-dimensional gravity coupled to higher-spin fields}},
  \href{https://doi.org/10.1007/JHEP11(2010)007}{\emph{JHEP} {\bfseries 11}
  (2010) 007} [\href{https://arxiv.org/abs/1008.4744}{{\ttfamily 1008.4744}}].

\bibitem{Gutperle:2011kf}
M.~Gutperle and P.~Kraus, \emph{{Higher Spin Black Holes}},
  \href{https://doi.org/10.1007/JHEP05(2011)022}{\emph{JHEP} {\bfseries 05}
  (2011) 022} [\href{https://arxiv.org/abs/1103.4304}{{\ttfamily 1103.4304}}].

\bibitem{Balachandran:1991dw}
A.~P. Balachandran, G.~Bimonte, K.~S. Gupta and A.~Stern, \emph{{Conformal edge
  currents in Chern-Simons theories}},
  \href{https://doi.org/10.1142/S0217751X92002106}{\emph{Int. J. Mod. Phys. A}
  {\bfseries 7} (1992) 4655}
  [\href{https://arxiv.org/abs/hep-th/9110072}{{\ttfamily hep-th/9110072}}].

\bibitem{Banados:1994tn}
M.~Banados, \emph{{Global charges in Chern-Simons field theory and the (2+1)
  black hole}}, \href{https://doi.org/10.1103/PhysRevD.52.5816}{\emph{Phys.
  Rev. D} {\bfseries 52} (1996) 5816}
  [\href{https://arxiv.org/abs/hep-th/9405171}{{\ttfamily hep-th/9405171}}].

\bibitem{Regge:1974zd}
T.~Regge and C.~Teitelboim, \emph{{Role of Surface Integrals in the Hamiltonian
  Formulation of General Relativity}},
  \href{https://doi.org/10.1016/0003-4916(74)90404-7}{\emph{Annals Phys.}
  {\bfseries 88} (1974) 286}.

\bibitem{Compere:2013bya}
G.~Comp\`ere, W.~Song and A.~Strominger, \emph{{New Boundary Conditions for
  AdS3}}, \href{https://doi.org/10.1007/JHEP05(2013)152}{\emph{JHEP} {\bfseries
  05} (2013) 152} [\href{https://arxiv.org/abs/1303.2662}{{\ttfamily
  1303.2662}}].

\bibitem{Compere:2015knw}
G.~Comp\`ere, P.~Mao, A.~Seraj and M.~M. Sheikh-Jabbari, \emph{{Symplectic and
  Killing symmetries of AdS$_{3}$ gravity: holographic vs boundary gravitons}},
  \href{https://doi.org/10.1007/JHEP01(2016)080}{\emph{JHEP} {\bfseries 01}
  (2016) 080} [\href{https://arxiv.org/abs/1511.06079}{{\ttfamily
  1511.06079}}].

\bibitem{Grumiller:2016pqb}
D.~Grumiller and M.~Riegler, \emph{{Most general AdS$_{3}$ boundary
  conditions}}, \href{https://doi.org/10.1007/JHEP10(2016)023}{\emph{JHEP}
  {\bfseries 10} (2016) 023}
  [\href{https://arxiv.org/abs/1608.01308}{{\ttfamily 1608.01308}}].

\bibitem{Afshar:2016wfy}
H.~Afshar, S.~Detournay, D.~Grumiller, W.~Merbis, A.~Perez, D.~Tempo et~al.,
  \emph{{Soft Heisenberg hair on black holes in three dimensions}},
  \href{https://doi.org/10.1103/PhysRevD.93.101503}{\emph{Phys. Rev. D}
  {\bfseries 93} (2016) 101503}
  [\href{https://arxiv.org/abs/1603.04824}{{\ttfamily 1603.04824}}].

\bibitem{Afshar:2016kjj}
H.~Afshar, D.~Grumiller, W.~Merbis, A.~Perez, D.~Tempo and R.~Troncoso,
  \emph{{Soft hairy horizons in three spacetime dimensions}},
  \href{https://doi.org/10.1103/PhysRevD.95.106005}{\emph{Phys. Rev. D}
  {\bfseries 95} (2017) 106005}
  [\href{https://arxiv.org/abs/1611.09783}{{\ttfamily 1611.09783}}].

\bibitem{Perez:2016vqo}
A.~P\'erez, D.~Tempo and R.~Troncoso, \emph{{Boundary conditions for General
  Relativity on AdS$_{3}$ and the KdV hierarchy}},
  \href{https://doi.org/10.1007/JHEP06(2016)103}{\emph{JHEP} {\bfseries 06}
  (2016) 103} [\href{https://arxiv.org/abs/1605.04490}{{\ttfamily
  1605.04490}}].

\bibitem{Ojeda:2019xih}
E.~Ojeda and A.~P\'erez, \emph{{Boundary conditions for General Relativity in
  three-dimensional spacetimes, integrable systems and the KdV/mKdV
  hierarchies}}, \href{https://doi.org/10.1007/JHEP08(2019)079}{\emph{JHEP}
  {\bfseries 08} (2019) 079}
  [\href{https://arxiv.org/abs/1906.11226}{{\ttfamily 1906.11226}}].

\bibitem{Sheikh-Jabbari:2016unm}
M.~M. Sheikh-Jabbari and H.~Yavartanoo, \emph{{On 3d bulk geometry of Virasoro
  coadjoint orbits: orbit invariant charges and Virasoro hair on locally
  AdS$_3$ geometries}},
  \href{https://doi.org/10.1140/epjc/s10052-016-4326-z}{\emph{Eur. Phys. J. C}
  {\bfseries 76} (2016) 493}
  [\href{https://arxiv.org/abs/1603.05272}{{\ttfamily 1603.05272}}].

\bibitem{Witten:1988hc}
E.~Witten, \emph{{(2+1)-Dimensional Gravity as an Exactly Soluble System}},
  \href{https://doi.org/10.1016/0550-3213(88)90143-5}{\emph{Nucl. Phys. B}
  {\bfseries 311} (1988) 46}.

\bibitem{Coussaert:1995zp}
O.~Coussaert, M.~Henneaux and P.~van Driel, \emph{{The Asymptotic dynamics of
  three-dimensional Einstein gravity with a negative cosmological constant}},
  \href{https://doi.org/10.1088/0264-9381/12/12/012}{\emph{Class. Quant. Grav.}
  {\bfseries 12} (1995) 2961}
  [\href{https://arxiv.org/abs/gr-qc/9506019}{{\ttfamily gr-qc/9506019}}].

\bibitem{Llabres:2019jtx}
E.~Llabr\'es, \emph{{General solutions in Chern-Simons gravity and $
  T\overline{T} $-deformations}},
  \href{https://doi.org/10.1007/JHEP01(2021)039}{\emph{JHEP} {\bfseries 01}
  (2021) 039} [\href{https://arxiv.org/abs/1912.13330}{{\ttfamily
  1912.13330}}].

\bibitem{Datta:2021kha}
S.~Datta and Y.~Jiang, \emph{{Characters of irrelevant deformations}},
  \href{https://doi.org/10.1007/JHEP07(2021)162}{\emph{JHEP} {\bfseries 07}
  (2021) 162} [\href{https://arxiv.org/abs/2104.00281}{{\ttfamily
  2104.00281}}].

\bibitem{Bunster:2014mua}
C.~Bunster, M.~Henneaux, A.~Perez, D.~Tempo and R.~Troncoso, \emph{{Generalized
  Black Holes in Three-dimensional Spacetime}},
  \href{https://doi.org/10.1007/JHEP05(2014)031}{\emph{JHEP} {\bfseries 05}
  (2014) 031} [\href{https://arxiv.org/abs/1404.3305}{{\ttfamily 1404.3305}}].

\end{thebibliography}
%%%%%%%%%%%%%%%%%%%%%%%%%%%%%%%%

\providecommand{\href}[2]{#2}\begingroup\raggedright\endgroup

\end{document}